\documentstyle [psfig,supertab] {l-aa}  
 
\def\apj{ApJ }
\def\apjs{ApJS }

\def\aj{AJ }
\def\aa{A\&A }
\def\aas{A\&AS }
\def\araa{ARA\&A }
\def\mnras{MNRAS }

\def\pasp{PASP }

\def\iau143{1991, in ``Wolf-Rayet stars and Interrelations with Other Massive Stars in 
	Galaxies'', IAU Symp.~143, Eds.~K.A.~Van der Hucht, B.~Hidayat, Kluwer}
\def\iau145{YEAR, in ``Evolution of Stars: the photospheric abundance 
	connection'', IAU Symp.~145, Eds.~G.~Michaud, A.~Tutukov, Kluwer}
\def\iau163{1994, in ``Wolf-Rayet stars: Colliding winds, binaries, evolution'',
	Eds.~K.A.~van der Hucht, P.M.~Williams, Kluver Academic Publ., Dordrecht}

\def\etal{et al.~}

\def\pder#1#2{\ifmmode  {\frac{\partial #1}{\partial #2}}\fi}
\def\lder#1{\ifmmode  {\frac{D #1}{Dt}}\fi}

\def\masl{\ifmmode  {\rm M_{\sun}yr^{-1}} \else ${\rm M_{\sun}yr^{-1}}$\fi}

\def\mdot{\ifmmode  \dot{M} \else $\dot{M}$\fi}
\def\msun{\ifmmode M_{\odot} \else $M_{\odot}$\fi}
\def\vinf{\ifmmode v_{\infty} \else $v_{\infty}$\fi}
\def\teff{\ifmmode T_{\rm eff} \else $T_{\rm eff}$\fi}
\def\logg{\ifmmode \log g \else $\log g$\fi}
\def\loggeff{\ifmmode \log g_{\rm eff} \else $\log g_{\rm eff}$\fi}
\def\rstar{\ifmmode R_{\star} \else $R_{\star}$\fi}
\def\lstar{\ifmmode L_{\star} \else $L_{\star}$\fi}
\def\mstar{\ifmmode M_{\star} \else $M_{\star}$\fi}
\def\rsun{\ifmmode R_{\odot} \else $R_{\odot}$\fi}
\def\lsun{\ifmmode L_{\odot} \else $L_{\odot}$\fi}
\def\12c16o{$^{12}{\rm C}\left(\alpha,\gamma\right)^{16}{\rm O}$}
\def\kms{\ifmmode {\rm km \;s^{-1}} \else $\rm km \;s^{-1}$\fi}
\def\nlte{non--LTE}
\def\lb{line blanketing}

\def\taustar{\ifmmode \tau_{\star} \else $\tau_{\star}$\fi}
\def\tauross{\ifmmode \tau_{\rm Ross} \else $\tau_{\rm Ross}$\fi}
\def\teffstar{\ifmmode{T^{\star}_{\rm eff}} \else $T^{\star}_{\rm eff}$\fi}

\def\ang{\ifmmode {\rm \AA} \else $\rm \AA$\fi}  % Angstroem
\def\hi{H~{\sc i}}
\def\hii{H~{\sc ii}}
\def\hei{He~{\sc i}}
\def\heii{He~{\sc ii}}

\def\he0{\ifmmode {\rm He^{\circ}} \else $\rm He^{\circ}$\fi}
\def\hep{\ifmmode {\rm He^+} \else $\rm He^{+}$\fi}
\def\hepp{\ifmmode {\rm He^{2+}} \else $\rm He^{2+}$\fi}
\def\halpha{\ifmmode {\rm H{\alpha}} \else $\rm H{\alpha}$\fi}
\def\hgamma{\ifmmode {\rm H{\gamma}} \else $\rm H{\gamma}$\fi}
\input rotate
\def\rotfig#1#2#3{\epsfxsize=#3cm \newbox\rotbox \setbox\rotbox=\hbox{\epsfbox{#2}}\centering\noindent\rotr\rotbox    \vspace{-0.4cm}}
\begin{document}

\thesaurus{07(08.01.3; 08.13.2; 08.05.1; 13.25.5; 09.08.1)}
 
\title{Combined stellar structure and atmosphere models for massive stars}
\subtitle{III. Spectral evolution and revised ionising fluxes of O3--B0 stars}

\author{Daniel Schaerer\inst{1,}\thanks{{\it Present address:\/} 
Space Telescope Science Institute,
3700 San Martin Drive, Baltimore, MD 21218, USA (schaerer@stsci.edu)}
\and Alex de Koter\inst{2}}
\offprints{D. Schaerer }

\institute{Geneva Observatory, CH-1290 Sauverny, Switzerland  \and
	NASA/GSFC, Advanced Computer Concepts, Code 681, Greenbelt, MD 20771}

\date{Received 17 May 1996 / Accepted 25 October 1996}
\maketitle
\markboth{D.~Schaerer \& A.~de Koter: Spectral evolution and revised 
	ionising fluxes of O3--B0 stars}
	{D.~Schaerer \& A.~de Koter: Spectral evolution and revised 
	ionising fluxes of O3--B0 stars}

\begin{abstract}
We provide an extensive set of theoretical spectral energy distributions
of massive stars derived from our ``combined stellar structure and 
atmosphere models''. The calculations cover the entire 
main sequence evolution for initial masses $M_{\rm i}=$ 20 -- 120 \msun, 
corresponding to O3--B0 stars of all luminosity classes.

We predict detailed line blanketed UV spectra along the main
sequence evolution.
The major result is a systematic study of ionizing fluxes 
covering the entire parameter space of O and early B stars.
We demonstrate the importance of accounting simultaneously for 
\nlte\ effects, line blanketing and stellar winds to obtain
an accurate description of the spectra of these stars shortward of the 
Lyman limit.
The main results from our spectra are the following: 
\begin{itemize}
\item[$\diamond$]   The flux in the \heii\ continuum is increased 
	by 2 to 3 (3 to 6) orders of magnitudes compared to predictions from 
	plane parallel \nlte\ (LTE) model atmospheres. This reconfirms
        the work of Gabler \etal (1989).
\item[$\diamond$] 	
	The flux in the \hei\ continuum is known to be increased due
	\nlte\ effects. However, we find that it is also
	influenced by wind effects as was previously found by 
        Najarro \etal (1996) and Schaerer \etal (1996b).
	The combined effect of a mass outflow and \lb\ 
	leads to a flatter energy distribution in the 
	\hei\ continuum, which confirms the results 
	of Sellmaier \etal (1996) for a wider range of stellar
 	parameters.
\item[$\diamond$] 
	The flux in the Lyman continuum is also modified due to 
        \lb\ and the presence of a stellar
	winds, although to a lesser degree than the spectrum at higher
	energies.
\end{itemize}

We derive revised ionizing fluxes for O3 to 
B0 stars based on the recent temperature and gravity calibrations of 
Vacca \etal (1996).
The total number of Lyman continuum photons is found to be
slightly lower than previous derivations.
For most cases the differences are less than $\sim$ 20 \%.
Due to the increased flux in the \hei\ continuum the hardness ratio 
of the \hei\ to H continuum is increased by $\sim$ 1.6 to
$\sim$ 2.5 depending on spectral type and luminosity class.

In the view of recent EUV and X-ray observations, a critical 
discussion of current model assumptions (including our own) 
shows that for stars of spectral types later than approximately B0,
which have relatively weak stellar winds, reliable predictions of 
ionizing fluxes are not yet possible. We identify the most likely
physical reasons for this finding.

\keywords{ stars: atmospheres  -- stars: mass--loss -- stars: early--type  --  
ultraviolet: stars -- \hii\ regions}

\end{abstract}

%%%%%%%%%%%%%%%%%%%%%%%%%%%%%%%%%%%%%%%%%%%%%%%%%%%%%%%%%%%%%%%%%%%%%%%%
\section{Introduction}
\label{s_intro}
The recent combined stellar structure and atmosphere ({\em CoStar}) models
for massive stars of Schaerer \etal (1996a, b, hereafter Paper I and II) 
consistently treat the stellar interior, photosphere and wind
using up-to-date input physics. 
An immediate advantage of this approach is that we predict 
the emergent spectral energy distribution along the evolutionary paths,
which provides a large number of observable quantities ranging from the
extreme ultraviolet (EUV) to the infrared (IR).
Of particular interest is the spectral range shortward of the Lyman
limit, where the bulk of the bolometric luminosity of hot star is emitted.
This wavelength range has been recently observed in early-type stars
by Hoare et al. (1993) and Cassinelli et al. (1995, 1996). 

From both theoretical and observational results,
it has become clear that the flux shortward of the Lyman edge 
is not only affected by \lb\ and \nlte\ effects but is also influenced by the 
presence of a stellar wind. The presence of a stellar wind may considerably 
change the formation of the 
flux in the Lyman continuum up to high frequencies. This was first pointed out by
Gabler \etal (1989), who showed that a wind can cause a significant depopulation
of the \heii\ ground state.
As a consequence, models accounting for stellar winds
lead to \hep\ ionizing fluxes which are 
$\sim$ 2--3 orders of magnitudes larger than the values predicted from \nlte\
plane parallel 
models (see Gabler \etal 1992, Paper II). With respect to
the widely used LTE models of Kurucz (1991) the increase is
3--6 orders of magnitude!
More recently evidence has been presented that the flux at longer wavelengths,
i.e.~in the $\rm He^{\circ}$ continuum (at $\lambda < 504$ \ang) and even in 
the Lyman continuum
(at $\lambda < 912$ \ang), can also be affected by the presence of a stellar wind
(Paper II, Najarro \etal 1996). As in the case of the \hep\ ionizing flux 
this occurs through a depopulation of the corresponding ground states.

The above illustrates the importance of treating the stellar wind when
predicting ionizing fluxes. The first results of these theoretical studies
have already had important consequences for the interpretation of observations:
\begin{itemize}
\item The strong increase of \hep\ ionizing photons in
the models of Gabler \etal (1991) leads to an important reduction of the
Zanstra--discrepancy in central stars of planetary nebul\ae.

\item Nebular calculations using emergent fluxes from wind models yields an
      improved match to
the observed ionization structure of \hii\ regions,
thereby resolving the so-called [Ne~{\sc iii}] problem (Sellmaier \etal 1996, 
Rubin \etal 1995 and references therein).
\item {\sc EUVE} observations of the Lyman continuum flux of the 
B2II giant $\epsilon$ CMa by Cassinelli \etal (1995) have provided a first {\em direct 
comparison} with model atmospheres. 
Surprisingly the observations show a flux significantly larger than predicted from both
LTE and \nlte\ plane parallel models. 
Among several alternative explanations for this failure of plane parallel atmosphere
models (Hubeny \& Lanz 1996), Najarro \etal (1996) suggest that the discrepancy 
of the Lyman continuum flux may be reduced if one accounts for the weak stellar 
wind of $\epsilon$ CMa. 
%(\mdot $\sim$ $10^{-8}$ $\msun yr^{-1}$).
\end{itemize}

The above findings
clearly point out the necessity to improve our understanding
of the ionizing fluxes of OB-type stars.
While the first studies have concentrated on individual objects and very few 
models are available yet, our study covers the entire main sequence evolution
between 20 and 120 \msun, i.e. spectral types O3 to early-B, and provides predictions
from elaborate \nlte\ calculations including \lb\ and stellar winds.
This will allow to work out the consequences of revised ionizing fluxes
for a large number of systems, including the galactic ISM, \hii\ regions and
starbursts.
Such direct or indirect comparisons with observations will, in turn, be of great 
value for testing our predictions and improve the reliability of model atmospheres
for hot stars.

The remainder of the paper is structured as follows:
Our method and the calculated model set are described in Sect.~2. 
Predicted UV line blanketed spectra are presented in Sect.~3. 
The main presentation and discussion of the ionizing fluxes takes up 
Sect.~4. Section 5 contains our revised calibration of ionizing fluxes
and contains comparisons to previous work.
In Sect.~6 we discuss uncertainties of the present models
and point out future improvements. The main results are 
summarised in Sect.~7.

%%%%%%%%%%%%%%%%%%%%%%%%%%%%%%%%%%%%%%%%%%%%%%%%%%%%%%%%%%%%%%%%%%%%%%%%
\section{Model calculations}
\label{s_calculations}

\subsection{Input physics and method}
\label{s_input}
A detailed description of our so-called {\em CoStar} models and the input 
physics adopted for the calculations is given in Paper I of this series.
Here, we will only briefly summarise the most important characteristics
of our models.

The entire star, comprising the stellar interior and a spherically expanding
atmosphere including the stellar wind, is treated consistently.
The interior is modelled with the Geneva stellar evolution code using the
same input physics (reaction rates, opacities etc.) as in the latest grid 
calculations of Meynet \etal (1994).
The atmosphere is modelled using the {\sc isa-wind} code of de Koter \etal 
(1993, 1996b).
Outer boundary conditions 
for the stellar interior calculations
are given by the atmospheric 
structure. 
%including the photosphere and the stellar wind. % (see Paper I). 
Basically the atmosphere is characterised by two parts: the subsonic regime 
with an extended photosphere and the wind, where the flow is accelerated to 
the terminal flow velocity \vinf. For the photospheric part we solve the 
stationary momentum equation taking into account gas and radiation pressure.
The subsonic part is smoothly connected with a wind structure described by
the usual ``$\beta$-law'' (see Paper I).
The temperature structure is given by radiative equilibrium in an extended
grey atmosphere following Lucy (1971) and Wessolowski \etal (1988).
In the final step, a consistent solution is constructed, embracing 
both the stellar interior and the atmosphere.  
In addition to the usual 
predictions from evolutionary models, {\em CoStar} models also provide the detailed 
emergent fluxes along the evolutionary paths.

The adopted mass loss rate and the additional parameters required to 
describe the wind structure are taken as in Paper I:
\begin{itemize}
\item Mass loss rates are adopted as in Meynet \etal (1994). This means
that for population I stars throughout the HR diagram we use the mass loss
rates given by de Jager et al. (1988), enhanced by a factor of two.
Justifications for this choice are given by Meynet \etal (1994) and
Maeder \& Meynet (1994). For non-solar metallicities \mdot\ was scaled
with $\left(Z/Z_\odot \right)^\zeta$, where $Z_\odot=0.020$. Consistent 
with our previous grid calculations an exponent $\zeta=0.5$ was taken 
as indicated by wind models (Kudritzki \etal 1987, 1991).

\item The terminal velocities \vinf\ as a function of metallicity are 
from wind models of Leitherer \etal (1992). 
Comparisons of our adopted terminal velocities with observations of 
population I stars have been discussed in Paper I.
% NOTE: Z dependence from LRD is compatible with KPP !

\item For the rate of acceleration of the supersonic flow 
we take $\beta=0.8$ following theoretical 
predictions of Friend \& Abbott (1986) and Pauldrach \etal (1986). 
These predictions are in good agreement with 
observations of O stars by Groenewegen \& Lamers (1991).
\end{itemize}

The {\sc isa-wind} \nlte\ radiation transfer calculations, which yield the detailed
spectral evolution use the atmospheric structure from the {\em CoStar} model 
summarised above.
In {\sc isa-wind}, the line transfer problem is treated using the Sobolev approximation, 
including the effects of the diffuse radiation field, and the continuous opacity 
inside the line resonance zone (de Koter 1993, de Koter \etal 1993).
Line blanketing is included 
following the opacity sampling technique introduced by Schmutz (1991). 
The method involves a Monte Carlo radiation transfer calculation including 
the most important spectral lines of all elements up to zinc.
We want to make clear that our models are not fully ``line blanketed''
in the context established in photospheric models, i.e.~a
fully consistent treatment of the effects of the presence
of lines on the atmospheric structure and emergent spectrum.
However, we opted to use this term rather than the term
``line blocked'' because the latter would go by the fact that
we {\em do} treat the redistribution of flux and -- although in
an approximate way -- the effect of blocking on the
temperature structure.
The ionization and excitation of the metals is treated as in Schaerer \& Schmutz
(1994a,b) to which the reader is referred for a detailed description of the entire 
procedure. 

The input physics for the atmospheric structure calculations
consists of atomic data for the elements explicitly included in the 
\nlte\ model. In the present work hydrogen and helium are treated 
as summarised in Paper I.
The H, He, C, N, and O composition of the atmosphere is that corresponding 
to the outermost layer of the interior model.
For the metals included in the line blanketed atmosphere, the abundances
of Anders \& Grevesse (1989) have been adopted.

The domain where our models are applicable is limited 
to relatively strong winds because of several simplifying assumptions 
made in the calculations (see also Sect.~\ref{s_improve} for
a critical discussion):
\begin{itemize}
\item[$\diamond$] 
The Sobolev approximation yields
good agreement with comoving frame calculations for O and WR
stars (de Koter \etal 1993). However, for weaker winds differences 
in the level populations will progressively affect the predicted 
continuum fluxes in particular shortward of the Lyman edge.

\item[$\diamond$] 
Presently our calculations neglect line broadening, yielding 
only a poor treatment of photospheric lines. This results in an 
underestimate of blanketing in the photosphere. 
For the early spectral types and/or strong winds this approximation
should not be crucial,
since in these cases photospheric lines are both weaker and less numerous,
and wind effects play a very important role in establishing the 
equilibrium population.
\item[$\diamond$] 
The temperature structure, which is derived from radiative 
equilibrium in an extended grey atmosphere, includes line blanketing
only in an approximate way. Even with the improved treatment of
Schaerer \& Schmutz (1994a), the determination of the temperature structure
in the photosphere-wind transition zone remains somewhat uncertain.
In the case that wind effects dominate, this effect should not be
of importance.
\item[$\diamond$] 
We neglect X-rays which, for stars with relatively weak
winds, can drastically alter the ionization structure (MacFarlane \etal 
1994) and might also provide an additional heating mechanism. 
\end{itemize}

Despite these uncertainties, we will argue below that the O stars covered in the 
present paper 
%($M_{rm ini} \ga$ 20 \msun, roughly corresponding to all O stars) 
should be quite adequately treated with our techniques.
However, we do note that because of
the above mentioned points and because of other indications
(see Sects.~\ref{s_euve},\ref{s_improve}), for B stars reliable 
predictions of the ionizing fluxes are not yet possible.
Future improvements will be necessary to extend the range 
of validity of the models.

\begin{figure}[htb]
%\centerline{\psfig{figure=figs/hr_logg_models.eps,height=8.8cm}}
\centerline{\psfig{figure=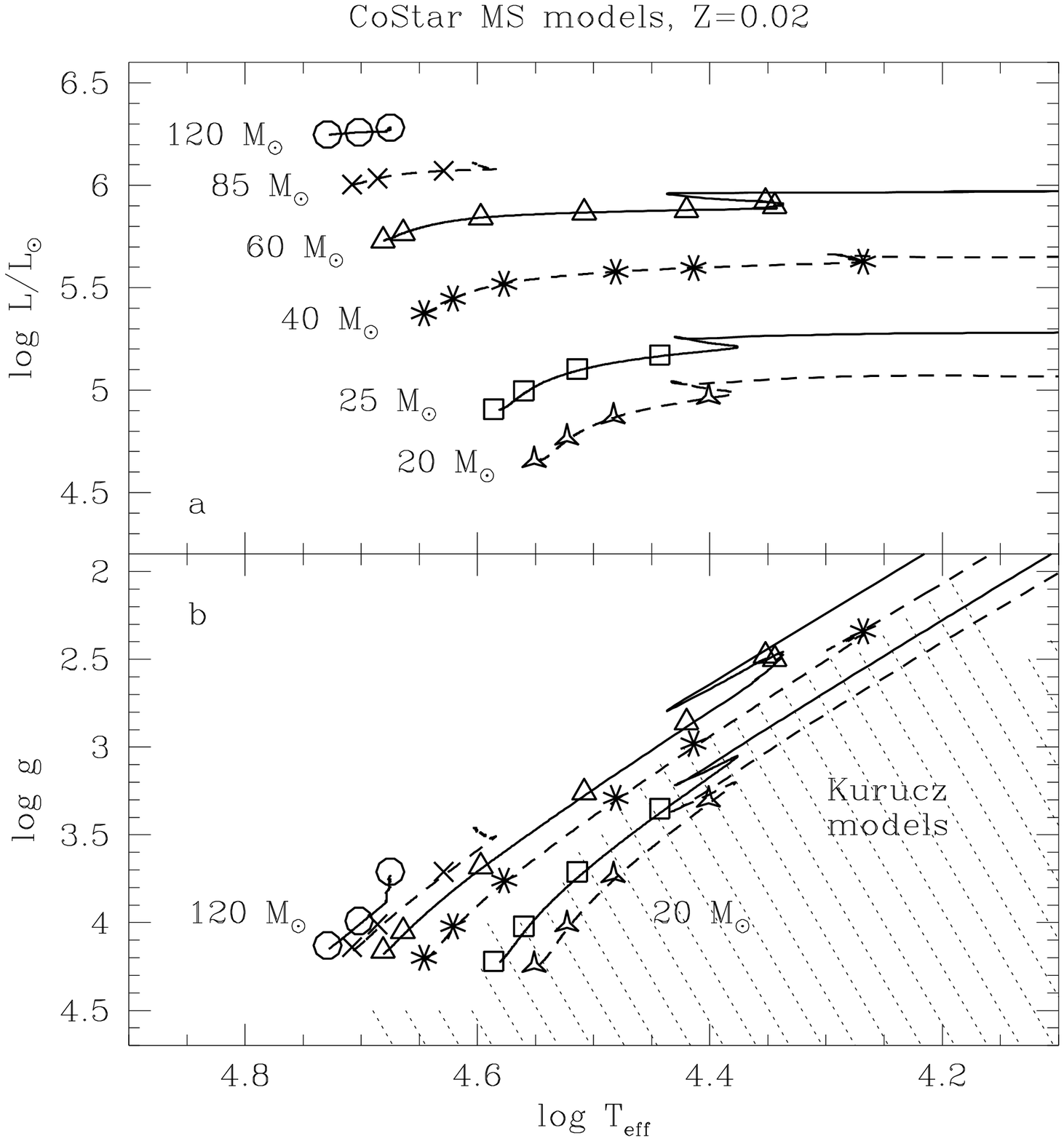,height=8.8cm}}
%\picplace{8.8cm}
\caption{{\bf a} HR--diagram covering the MS phases for all initial 
masses. The WR stage during the H--burning phase of the 85 and 120 \msun
models are not included. Symbols (circles, crosses etc.) denote the selected
models describing the spectral evolution (see Table \protect\ref{ta_params}).
{\bf b} $\logg$--$\log\teff$ diagram corresponding to {\bf a}. The hatched
area shows the domain for which Kurucz atmosphere models are available}
\label{fig_hr_logg_models}
\end{figure}

\begin{table*}
\caption{Summary of selected models at Z=0.020: stellar parameters and approximate spectral classification}
\centerline{
\begin{tabular}{rrlllllrrrrll}
\\ \hline \\
model & age & $\frac{M}{\msun}$ & $\log \teff$ & $\log \frac{L}{\lsun}$  & \logg & $\frac{\rstar}{\rsun}$ 
	& $\log \mdot$ & \vinf\ & $n_{\rm H}$ & $\mu$  & SpType & SpType\\
 \# & [yr] & & [K] & & [${\rm cm\; s^{-2}}$] & & [$\msun {\rm yr^{-1}}$] &  [\kms] & & & (\teff--$M_{\rm bol}$) & (HRD)\\         
\\ \hline \\
\multicolumn{6}{l}{20 \msun\ track:} \\		% *** A models ***
A1 & 4.00 $10^4$        & 20.00 & 4.551 & 4.658 & 4.24 &   5.633 & -7.058 & 2890. & 0.90 &  0.634 & O8V  & B0V \\ % 20 k0011
A2 & 3.65 $10^6$	& 19.60 & 4.523 & 4.767 & 4.01 &   7.268 & -6.857 & 2544. & 0.90 &  0.634 & O9V  & B0V \\ % 21 k0151
A3 & 6.15 $10^6$	& 19.15 & 4.483 & 4.872 & 3.73 &   9.860 & -6.634 & 2202. & 0.90 &  0.634 & B0V  & B0IV \\ % 22 k0251
A4 & 7.79 $10^6$	& 18.65 & 4.401 & 4.966 & 3.30 &  16.034 & -6.384 & 1802. & 0.90 &  0.634 & B0.5II & B0.5II \\ % 23 k0351
\multicolumn{6}{l}{25 \msun\ track:} \\		% *** B models ***
B1 & 2.75 $10^4$        & 25.00 & 4.586 & 4.907 & 4.22 &   6.387 & -6.806 & 2897. & 0.90 &  0.634 & O7V   & O9.5V \\ % 24 i0006
B2 & 2.60 $10^6$        & 24.48 & 4.560 & 4.997 & 4.02 &   7.994 & -6.596 & 2589. & 0.90 &  0.634 & O8V   & O9V \\ % 25 i0101
B3 & 4.79 $10^6$        & 23.71 & 4.514 & 5.103 & 3.71 &  11.187 & -6.307 & 2208. & 0.90 &  0.634 & O9IV  & O8.5V \\ % 26 i0211
B4 & 5.89 $10^6$        & 22.99 & 4.443 & 5.171 & 3.35 &  16.710 & -6.066 & 1868. & 0.90 &  0.634 & B0II  & O9.5III \\ % 27 i0291
\multicolumn{6}{l}{40 \msun\ track:} \\		% *** C models ***
C1 & 4.04 $10^4$ & 39.98	& 4.646 & 5.376 & 4.20  &  8.326 & -6.463 & 2962. & 0.90 &  0.634 & O5V     & O7V \\ % 1
C2 & 1.50 $10^6$ & 39.25	& 4.621 & 5.447 & 4.02	& 10.152 & -6.170 & 2690. & 0.90 &  0.634 & O6V     & O7V \\ % 2	
C3 & 2.85 $10^6$ & 37.86	& 4.577 & 5.519 & 3.76 	& 13.457 & -5.809 & 2354. & 0.90 &  0.634 & O7IV    & O7IV \\ % 3 
C4 & 3.81 $10^6$ & 35.48	& 4.481 & 5.578 & 3.29	& 22.422 & -5.414 & 1893. & 0.90 &  0.634 & O9.5II  & O9.5I \\ % 4
C5 & 4.08 $10^6$ & 34.35	& 4.414 & 5.597 & 2.98	& 31.215 & -5.290 & 1660. & 0.90 &  0.634 & B0I     & O9.5I \\ % 5
C6 & 4.38 $10^6$ & 32.38	& 4.268 & 5.628 & 2.34	& 63.521 & -5.126 & 1269. & 0.90 &  0.634 & B2I     & O9.5I \\ % 6
\multicolumn{6}{l}{60 \msun\ track:} \\  	% *** D models ***
D1 & 8.72 $10^4$	& 59.96 & 4.681 &  5.731 & 4.16	& 10.663 & -6.090 & 3050. & 0.90 &  0.634 & O4V     & O5.5V \\ % 7 
D2 & 7.72 $10^5$	& 59.20	& 4.664 &  5.766 & 4.05	& 12.010 & -5.856 & 2883. & 0.90 &  0.634 & O4.5V   & O5.5V \\ % 8
D3 & 2.23 $10^6$	& 55.03	& 4.597 &  5.842 & 3.68	& 17.838 & -5.252 & 2381. & 0.90 &  0.634 & O6III   & O5.5IV \\ % 9
D4 & 2.76 $10^6$	& 50.60	& 4.508 &  5.867 & 3.26	& 27.672 & -4.931 & 1960. & 0.90 &  0.634 & O9 I    & O7I \\ % 10
D5 & 3.00 $10^6$	& 47.43	& 4.420 &  5.879 & 2.86	& 42.147 & -4.826 & 1646. & 0.90 &  0.634 & B0I     & O7I \\ % 11
D6 & 3.23 $10^6$	& 43.68	& 4.344 &  5.897 & 2.50	& 60.890 & -4.779 & 1391. & 0.87 &  0.666 & B0.5I   & O7.5I \\ % 12
D7 & 3.44 $10^6$	& 40.01	& 4.352 &  5.920 & 2.48	& 60.392 & -4.747 & 1319. & 0.80 &  0.737 & B0.5I   & O7I \\ % 13
\multicolumn{6}{l}{85 \msun\ track:} \\		% *** E models ***
E1 & 5.00 $10^4$	& 84.88	& 4.708 & 6.004 & 4.14 &  12.900 & -5.683 & 3182. & 0.90 &  0.634 & O3IV    & O4V \\ % 14
E2 & 7.10 $10^5$	& 82.84	& 4.686 & 6.034 & 4.01 &  14.780 & -5.379 & 2977. & 0.90 &  0.634 & O3.5IV  & O4.5IV \\ % 15
E3 & 1.66 $10^6$	& 75.58	& 4.629 & 6.071 & 3.71 &  20.017 & -4.879 & 2538. & 0.90 &  0.634 & O5III   & O4.5III \\ % 16 
\multicolumn{6}{l}{120 \msun\ track:} \\	% *** F models ***
F1 & 4.18 $10^4$       & 119.54 & 4.728 & 6.248 & 4.13 &  15.603 & -5.147 & 3340. & 0.90 &  0.634 & O3V     & O3IV \\ % 17 h0011 !
F2 & 6.26 $10^5$       & 116.48 & 4.702 & 6.275 & 4.00 &  17.789 & -4.853 & 3103. & 0.90 &  0.634 & O3III   & O3.5III \\ % 18 h0071
F3 & 2.12 $10^6$        & 80.43 & 4.675 & 6.282 & 3.71 &  20.681 & -4.572 & 2428. & 0.75 &  0.795 & O3I     & O3.5I \\ % 19 h0251
\\ \hline
\end{tabular}
}
\label{ta_params}
\end{table*}

\subsection{Selected models}
\label{s_models}
To provide a complete coverage of the entire main sequence (MS) 
evolution with our {\em CoStar} models we have calculated evolutionary
tracks for initial masses of 20, 25, 40, 60, 85 and 120 \msun\
at solar metallicity.
This paper includes the results from Papers 
I and II, which covered the range from 40 to 85 \msun. 
Additionally, we provide calculations for the entire data set
at low metallicities (Z=0.004). 
A high metallicity grid is in preparation.
The continuum spectral energy distributions from both the Z=0.020 and 0.004
model sets are available on request from the authors. They will also be
included in a recent CD-ROM distributed by the AAS (Leitherer 
\etal 1996).

\subsubsection{Solar metallicity: Z=0.020}
Figure \ref{fig_hr_logg_models} shows the evolutionary tracks at Z=0.020 in the
HR-diagram and the $\logg$--$\log\teff$ diagram. 
For each track, we have selected several models for which the stellar
parameters are summarized in Table \ref{ta_params}.
Along each MS track the models have been selected according to the following 
criteria if possible :
{\em 1)} ZAMS model, {\em 2)-5)} $\logg$ approximately 4., 3.7, 3.3, 3., and 
{\em 6)} maximum radius. One additional TAMS model is available for the 60 
\msun\ track.
The total of 27 models should provide a good coverage of the entire MS spectral
evolution.

The following entries are given in Table \ref{ta_params}:
Model number (column 1), age (2), present mass (3), effective 
temperature (4), luminosity (5), gravity (6), stellar radius
(7), mass loss rate (8), terminal velocity (9), number fraction
of hydrogen $n_{\rm H}$ normalised to $n_{\rm H}+n_{\rm He}=1.$ (10) 
and the mean molecular weight per free particle $\mu$ (11) used to 
determine the photospheric structure. The last two columns
give an approximate spectral classification, which has been 
obtained from a nearest neighbour search in the tables of Schmidt-Kaler 
(1982)\footnote{At this point, we adopt the Schmidt-Kaler classification instead
of the more recent one from Vacca \etal (1996) since the latter does
not cover the entire domain of our models.}
The assignment in column 12 uses the variables ($\teff,M_{\rm bol}$)
\footnote{We specify \teff\ in kK, which gives 
a reasonably large weight to the temperature.},  
while a nearest neighbour search in the HR-diagram yields the spectral types 
in column 13.

\subsubsection{Low metallicity calculations: Z=0.004}
\label{s_z004}
For the low metallicity models, 
we have chosen the following approach: 
instead of calculating full {\em CoStar} models we have calculated atmosphere 
models at the same position in HR-diagram, i.e.~adopting identical stellar
parameters as for Z=0.020. Only the wind parameters (\mdot, \vinf)
and the composition have been adapted to Z=0.004.
This procedure allows detailed comparisons between the emergent spectra
at different metallicities.

The following changes apply for Z=0.004 with respect to the
parameters given for Z=0.020 in Table \ref{ta_params}:
\mdot\ is reduced by a factor of 2.236, and \vinf\ is reduced by 1.233
to reflect the dependence expected from wind models (cf.~Sect.~\ref{s_input}).
In addition to accounting
for the abundance changes of 
metals, one also needs to modify the relative hydrogen to helium abundance.
Consistently with recent evolutionary grid calculations 
(Charbonnel \etal 1993, Meynet \etal 1994), we adopt an initial helium
content, $Y=Y_p +\left(\Delta Y/\Delta Z\right)Z$ with
a $\Delta Y/\Delta Z$ ratio of 3. (see Schaller \etal 1992).  
The resulting abundances for Z=0.004 are: $(X,Y)$=(0.744,0.252) in mass
fraction, which corresponds to H and He number abundances of 
$(n_{\rm H},n_{\rm He})$=(0.922,0.078)
%\footnote{Remember that $n_{\rm H}$ 
%and $n_{\rm He}$ are normalised to $n_{\rm H}+n_{\rm He}=1.$}.
These abundances apply to all Z=0.004 models since, in contrast to the solar
metallicity tracks, no surface He enrichment is expected in this case
(see Meynet \etal 1994).
Consequently the mean molecular weight $\mu$,
%(cf.~Table \ref{ta_params}),
which enters in the determination of the atmospheric structure in the
low velocity domain is given by $\mu=0.600$.

For solar metallicity we will discuss several results in
detail.
For the reduced metallicity, we will present the integrated ionizing fluxes at 
Z=0.004 and briefly summarise the influence of metallicity on the emergent spectra.

% BELONGS to next section !
\begin{figure*}[htb]
%\centerline{\psfig{figure=figs/costar20_uv.eps,height=18cm}}
\centerline{\psfig{figure=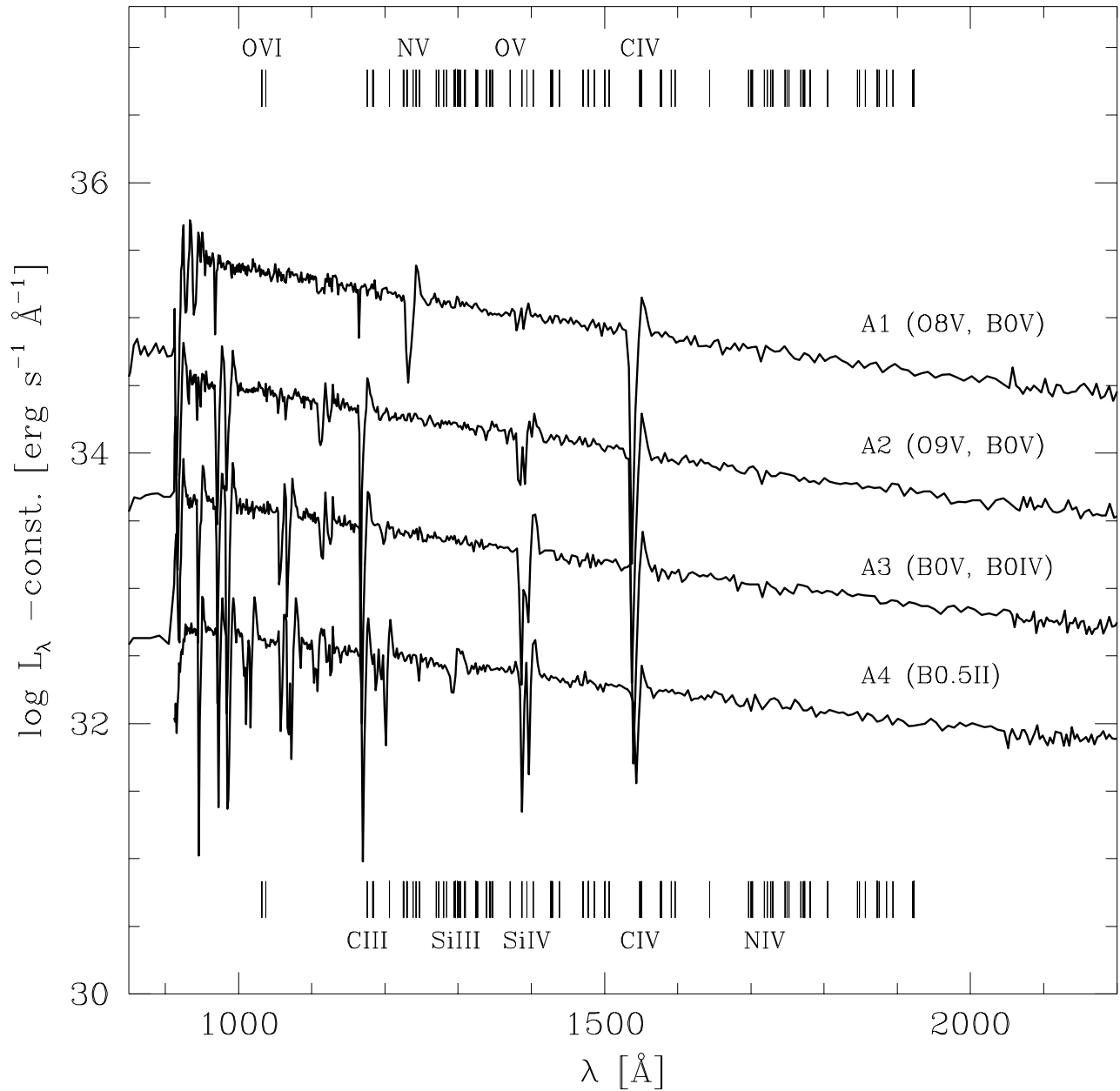,height=18cm}}
%\picplace{18cm}
\caption{Synthetic UV spectra showing the spectral evolution on the 20 \msun\ 
track. Plotted is the logarithm of the emergent luminosity.
Approximate spectral types from Table \protect\ref{ta_params} are given.
Starting with the second model, each spectrum has been 
shifted downwards by 0.7 dex with respect to the previous one, 
in order to allow a good comparison.
The marks on the top and bottom indicate the location of the CNO and Si
lines taken from the lists of Bruhweiler \etal (1981) and Dean \& Bruhweiler (1985).
The strongest CNO, and Si features are labeled}
\label{fig_costar20_uv}
\end{figure*}

\begin{figure*}[htb]
%\centerline{\psfig{figure=figs/costar25_uv.eps,height=18cm}}
\centerline{\psfig{figure=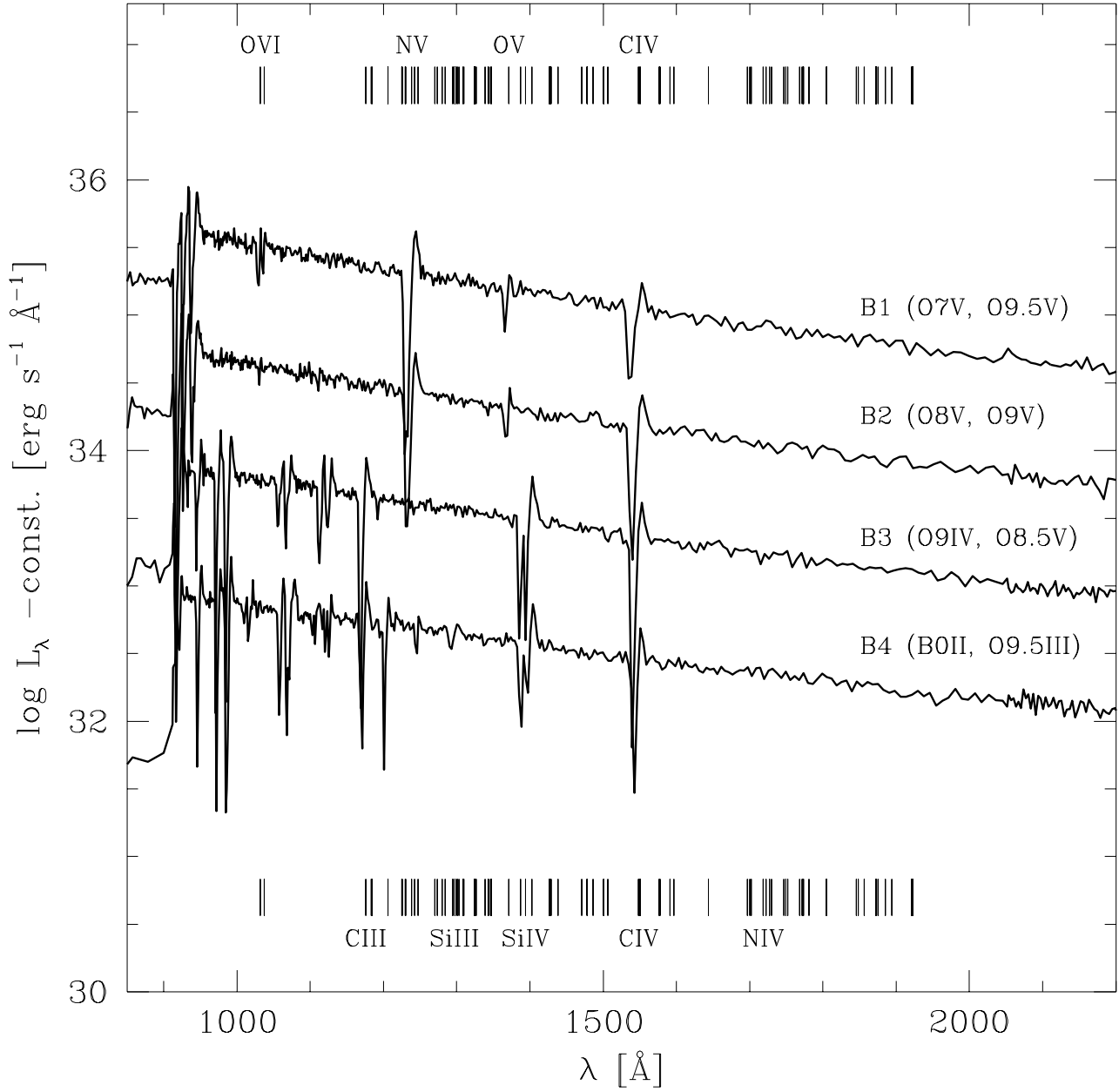,height=18cm}}
%\picplace{18cm}
\caption{Same as Fig.~\protect\ref{fig_costar20_uv} for the 25 \msun\ track
(models B1 to B4)}
\label{fig_costar25_uv}
\end{figure*}
\begin{figure*}[htb]
%\centerline{\psfig{figure=figs/costar120_uv.eps,height=18cm}}
\centerline{\psfig{figure=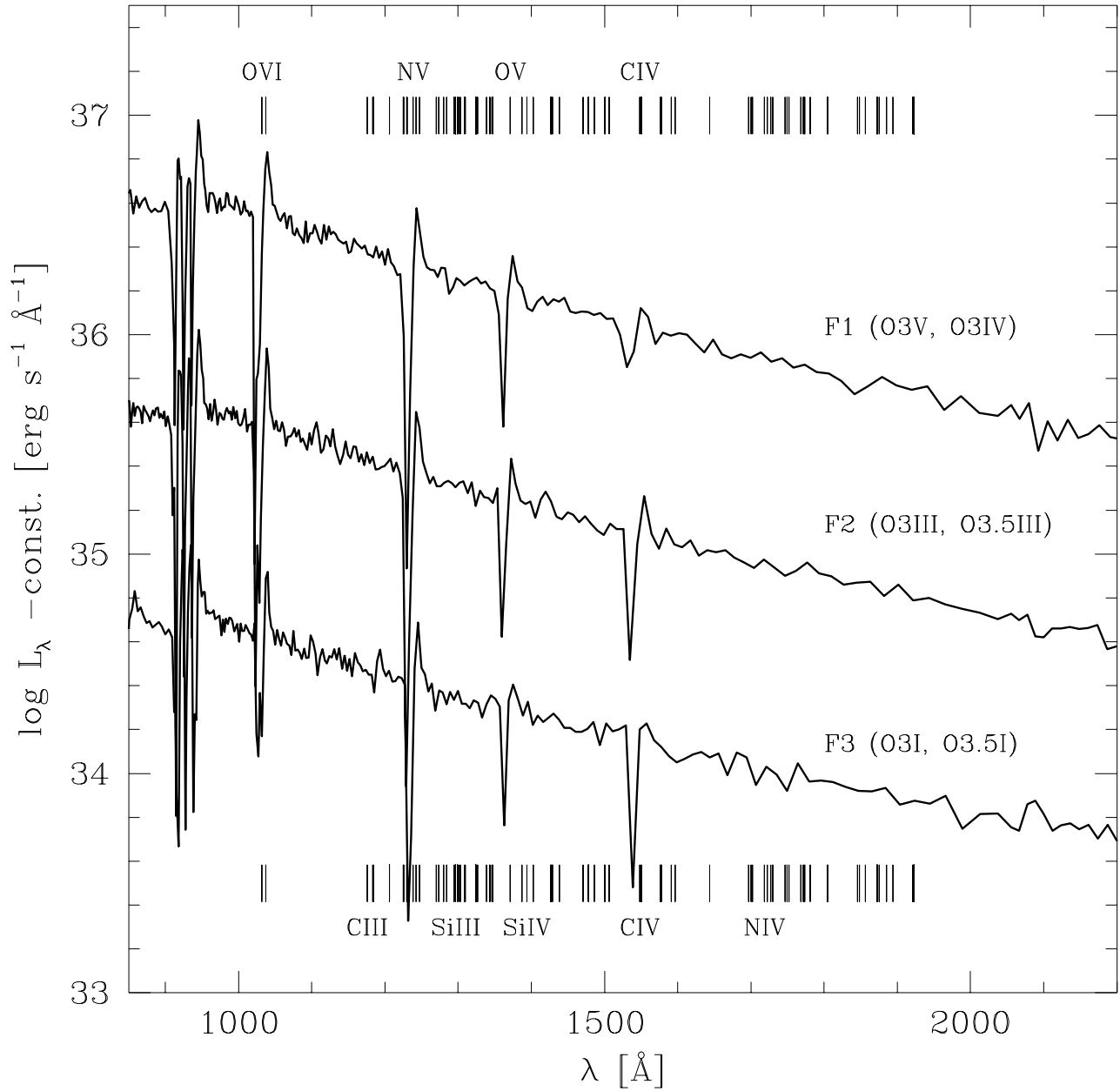,height=18cm}}
%\picplace{18cm}
\caption{Same as Fig.~\protect\ref{fig_costar20_uv} for the 120 \msun\ track
(models F1 to F3)}
\label{fig_costar120_uv}
\end{figure*}

\clearpage

%%%%%%%%%%%%%%%%%%%%%%%%%%%%%%%%%%%%%%%%%%%%%%%%%%%%%%%%%%%%%%%%%%%%%%%%
\section{Evolution of the UV line blanketed spectrum at Z=0.020}
\label{s_ms_blank}
Our line blanketed models allow us to predict a large number of 
detailed observable line features, which are well suited for comparison
with UV spectra. In Paper II we have presented the spectral evolution 
in the 850 to 2200 \ang\ range along MS tracks with initial masses between 40
and 85 \msun.  
To provide the complete set of UV spectra at solar metallicity we here
present the corresponding results for the additional tracks 
($M_{\rm i}$ = 20, 25 and 120 \msun). 

The synthetic UV spectra of the models given in Table\ref{ta_params} 
are plotted in Figs.~\ref{fig_costar20_uv} to \ref{fig_costar120_uv}.
Together with the plots from Paper II, these figures illustrate
the behaviour of the strongest UV features as a function of luminosity
and effective temperature.
The progressive behaviour of the strongest wind lines of CNO and Si 
agrees with the conclusions drawn in Paper II, where we refer to for more details.

{\em (a)} The predictions for the N~{\sc v} resonance line follows the observed decrease
          in line strength towards later spectral types. The spectra of the B-type stars show
          some tendency to produce less N~{\sc v} than observed, which reflects the problem
          of super-ionization.

{\em (b)} The strong luminosity dependence of the Si~{\sc iv} line is reproduced.

{\em (c)} While the C~{\sc iv} $\lambda$ 1550 doublet is predicted too weak for ZAMS models
	  with $M_{\rm i} \ga$ 40 \msun, we obtain a strong C~{\sc iv} P-Cygni line for less
	  massive ZAMS stars.
          Compared to observations (see Snow \etal 
	  1994) the C~{\sc iv} line of the 20 and 25 \msun\ ZAMS models is too strong. 
          Both results indicate that  
          the predicted carbon ionization balance is shifted towards too high
 	  ionization stages for a given temperature (see Paper~II).

{\em (d)} The predicted O~{\sc v} $\lambda$ 1371 line shows a strong 
	  P-Cygni profile for the hottest models, while it is mostly observed in absorption.
	  The reasons have been discussed in Paper II.

In Paper II we have performed several {\em quantitative} comparisons of metal line features.
In particular we have been able to reproduce the strong observed Fe~{\sc 1920} \ang\
feature in late O and early-B giants and supergiants. This feature was found to be a 
good temperature indicator for these stars.
The new tracks confirm this result. 

%%%%%%%%%%%%%%%%%%%%%%%%%%%%%%%%%%%%%%%%%%%%%%%%%%%%%%%%%%%%%%%%%%%%%%%%
\section{Ionizing spectrum}
\label{s_ionis}

In this section, we will compare our model predictions, which take into
account non-LTE effects, line blanketing and a stellar wind, with both
LTE and non-LTE plane parallel atmospheres. In this way, we can investigate
the effect of a stellar wind on the emerging EUV continuum flux. Before
doing so, we first briefly discuss the physics of the processes connected
with the presence of a stellar wind that are found to affect the flux 
distribution shortwards of the Lyman continuum edge. 
There are three categories of effects, wind velocity effects,
geometrical wind effects, and line blanketing effects.

%\paragraph{Wind velocity effects}
{\bf Wind velocity effects:}
Because of the velocity gradient in a stellar wind, the He~{\sc ii} ({\sc i}) 
groudstate may become depopulated. If this occurs in the 
region where the corresponding continuum is formed, this will lead to a 
decrease of the bound-free opacity and consequently to an increase 
of the flux at $\lambda < 228~(504)$ \AA.

The depopulation effect works as follows. We consider a point $r$ in the wind,
hereafter the local point. The velocity gradient in the wind flow 
allows helium resonance line photons to escape from $r$
to infinity, forcing this transition out of detailed balance. 
The $n=1$ level population is no longer
controlled by the local contribution to the mean intensity in the line, 
$\overline{J}$, i.e. by the local line
source function. Instead, $\overline{J}$ is dominated by the non-local
contribution which is proportional to the specific intensity at the
photosphere. Because the radiation temperature at the photosphere is
higher than the electron temperature $T(r)$ and because
the helium resonance transition is in the Wien part of the spectrum, this 
will lead to a large increase of radiation at the line frequency, 
depopulating the groundlevel (see Gabler et al.~1989). 

At a certain point in the wind, the depopulation reaches a minimum. Farther
out, the $n=1$ population will again increase because of the proportionality
of $\overline{J}$ to the dilution factor and because of an increased importance
of recombinations from higher levels (see Najarro et al.~1996).
We need to distinguish four cases:

{\em   (a-i)} The continuum is formed in the region below (i.e. at higher 
	      Rosseland optical depth) that of the regime of strong depopulation 
	      of the groudstate.
              This would correspond to stars with low density winds. If the mass 
	      loss is so small that the continuum is
              formed in the photosphere, it will be likely that no significant 
	      changes from plane parallel models have occurred, consequently, 
	      the continuum flux will not differ much from a model without a stellar wind.

{\em (a-ii)}  As in the first case, the mass loss is sufficiently low that the
              groundlevel does not suffer from the depopulation effect described above.
              What defines this case is that transitions between high lying levels 
              become optically thin. Subsequent electron cascading causes the ground 
              state population to increase.
              Consequently, the continuum flux will decrease somewhat relative to
              a model without a stellar wind.
              This case occurs in the \hei\ continuum of most of the models considered
              in this work. In particular, the described effect 
              causes a {\em flattening} of the spectrum in the \hei\ 
              continuum, since the continuum at relatively short 
              wavelengths is formed at relatively large Rosseland optical depth,
              where it is less sensitive to the described effect.

{\em (a-iii)} The continuum is formed in the geometrical region corresponding to
              that of strong depopulation. The continuum flux will be increased 
              significantly relative to a model without a stellar wind.

{\em (a-iv)}  The continuum is formed in the region outside of that of the 
	      depopulation
              regime. This would correspond to models with a very dense wind.
              An increased flux relative to a model without a stellar wind
              may still be present, but the flux will not be as high as in the 
	      preceding case.  

Case {\em  (a-ii)} typically occurs in the \hei\ continuum of most O-type
models (cf.~also Sellmaier \etal 1996). 
Case {\em (a-iii)} dominates the He~{\sc ii} continuum of O-type stars.
It may, however, also occur in the He~{\sc i} continuum of B-type stars (Najarro 
et al.~1996). 

%\paragraph{Geometrical wind effects}
{\bf Geometrical wind effects:}
{\em  (b)} 
Related to the above effects is the possibility that continuum forming layers
may be located outside of the photosphere. 
In this case the flux is determined by the competing effects of 
{\em (b-i))} the increase of the emitting surface yielding a larger flux, 
and {\em (b-ii)} a drop in the local source function, at the emitting surface,
due to the temperature decrease, which reduces the emergent flux.
The geometrical effects mainly occur in the \heii\ continuum, where
the opacity is the largest.

%\paragraph{Line blanketing effects}
{\bf Line blanketing effects:}
Line blanketing may either cause an increase or a decrease of the flux, 
because two competing effects play a role:

{\em  (c-i)} The presence of many lines in the photosphere causes a redistribution 
             of photons from the radiation field at $\lambda < 228~(504)$ \AA\ 
             towards photons of longer wavelength.
             This blocking of photospheric flux causes a decrease in the helium 
	     ionization, increasing the He~{\sc ii} ({\sc i}) groundlevel population.
             Both the photospheric blocking and the increased He~{\sc ii} ({\sc 
	     i}) groundstate population cause a decrease of the flux at 
	     $\lambda < 228~(504)$ \AA.

{\em (c-ii)} As first shown by Schaerer \& Schmutz (1994a),
	     the presence of many resonance lines in the stellar wind causes an 
	     increase of the isotropy of the radiation field. This diffuse 
	     radiation is essentially the result of (multiple) photon scattering 
	     and occurs most effectively in wavelength
             regions where the line density is so large that the lines overlap. 
	     In these regions the star effectively shows a larger geometrical dilution 
	     factor compared to the case of no line blanketing. The associated 
	     increase in mean intensity yields a higher ionization, consequently 
	     a decrease in the groundstate population of He~{\sc ii} ({\sc i}) 
	     which causes an increase of the flux at $\lambda < 228~(504)$ \AA.

The first effect is usually more important in H~{\sc i} and He~{\sc i} continua
and also dominates in the He~{\sc ii} continuum of stars with low density winds.
The last effect 
dominates in the He~{\sc ii} continuum of stars with high
density winds.

% 120 Msun model, logg=4. !
\begin{figure}[htb]
%\centerline{\psfig{figure=figs/compare_18ryd.eps,height=8.8cm}}
\centerline{\psfig{figure=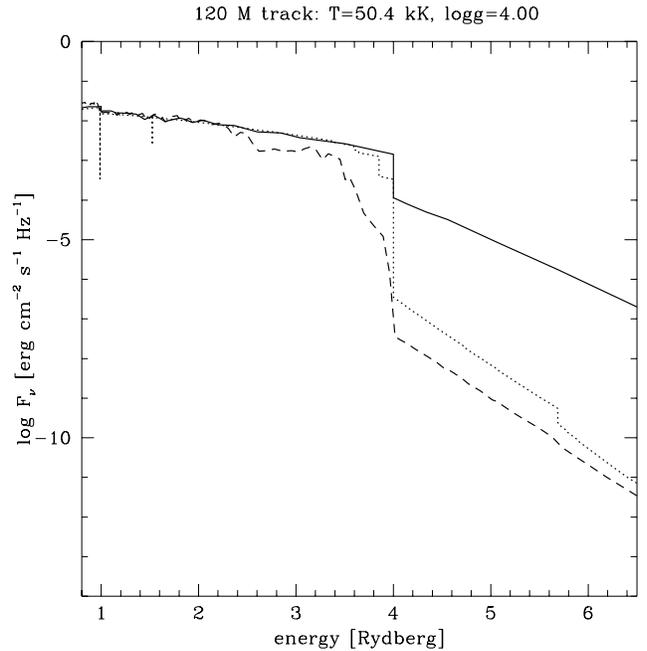,height=8.8cm}}
%\picplace{8.8cm}
\caption{Comparison of emergent EUV fluxes from a {\em CoStar} model
(solid line), a plane parallel \nlte\ model of Kunze (1994,
dotted line) and a plane parallel LTE model of Kurucz (1991, 
dashed line). Plotted is the astrophysical flux as a function 
of the energy.
The models are for a 120 \msun\ track dwarf model with parameters:
{\em CoStar}: model F2, 
%(cf.~Tab.~\protect\ref{ta_params}), 
Kunze: $(\teff,\logg)$ = (50 kK, 4.0) and
Kurucz: $(\teff,\logg)$ = (50 kK, 5.0)}
\label{fig_compare_18ryd}
\end{figure}

% 60 Msun models !
\begin{figure}[htb]
%\centerline{\psfig{figure=figs/compare_8ryd.eps,height=8.8cm}}
\centerline{\psfig{figure=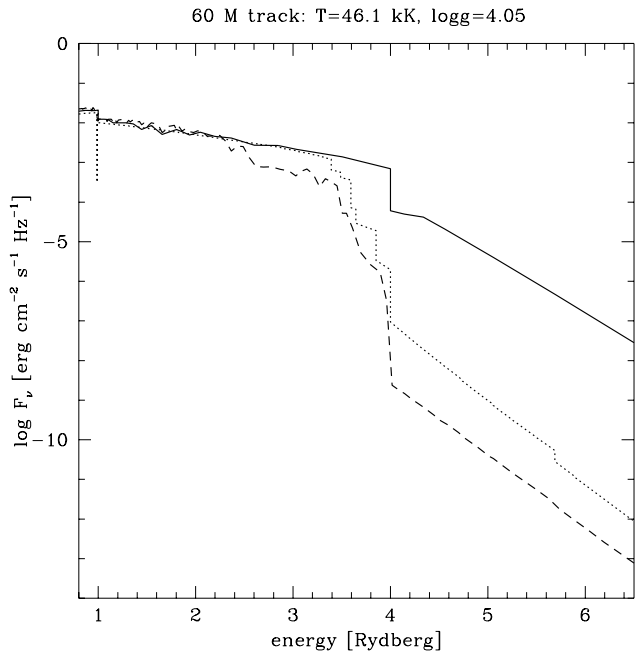,height=8.8cm}}
%\picplace{8.8cm}
\caption{Same as Fig.~\protect\ref{fig_compare_18ryd} for
a 60 \msun\ dwarf.
{\em CoStar}: model D2 (solid), 
Kunze (dotted):  $(\teff,\logg)$ = (45 kK, 4.0) and
Kurucz (dashed): $(\teff,\logg)$ = (45 kK, 5.0)}
\label{fig_compare_8ryd}
\end{figure}
\begin{figure}[htb]
%\centerline{\psfig{figure=figs/compare_10ryd.eps,height=8.8cm}}
\centerline{\psfig{figure=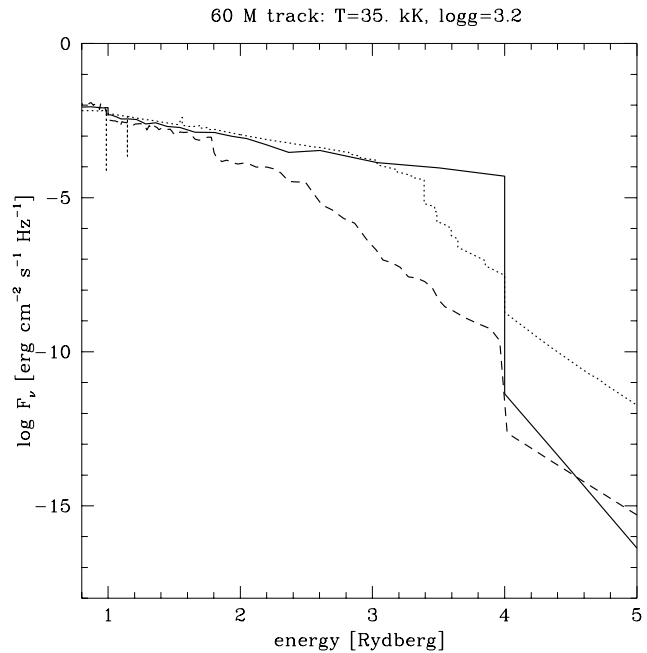,height=8.8cm}}
%\picplace{8.8cm}
\caption{Same as Fig.~\protect\ref{fig_compare_18ryd} for
a 60 \msun\ supergiant.
{\em CoStar}: model D4 (solid line). Parameters are as in 
Table.~\protect\ref{ta_params}, except for $(\teff,\logg)$ = (35 kK, 3.2), 
Kunze (dotted):  $(\teff,\logg)$ = (35 kK, 3.2), and
Kurucz (dashed): $(\teff,\logg)$ = (35 kK, 4.0)}
\label{fig_compare_10ryd}
\end{figure}

\subsection{Comparisons of CoStar and plane parallel models}
\label{s_costar}
We first compare the EUV spectral range of {\em CoStar} models with
plane parallel models.
We limit our comparisons to models which include \lb. 
For the plane parallel models we adopt the widely used LTE models 
of Kurucz (1991, ATLAS9) and the recent \nlte\ models from Kunze 
(1994, cf.~Kunze \etal 1992). The latter provide an extensive grid, 
covering essentially the same $\logg$--$\log \teff$ parameter 
space as the present {\em CoStar} calculations. 
The Kunze models include 9 of the most abundant elements in \nlte\ 
(H, He, C, N, O, Ne, Mg, Al, Si, S and Ar).
Grids for plane parallel \nlte\ models, 
which also include iron are not yet available 
(e.g.~Dreizler \& Werner 1993, Hubeny \& Lanz 1995).

% 20 Msun models !
\begin{figure}[htb]
%\centerline{\psfig{figure=figs/compare_20ryd.eps,height=8.8cm}}
\centerline{\psfig{figure=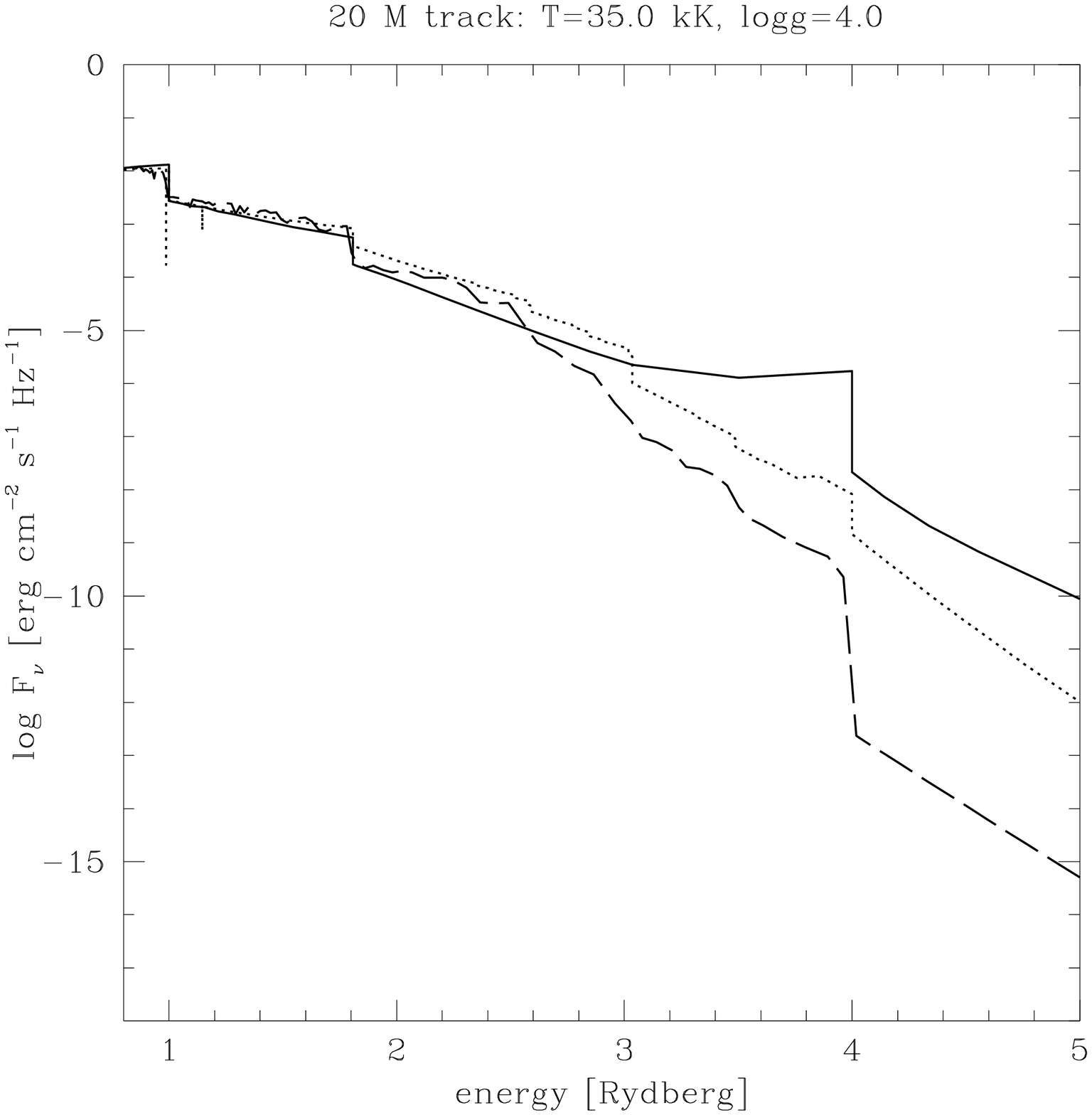,height=8.8cm}}
%\picplace{8.8cm}
\caption{Same as Fig.~\protect\ref{fig_compare_18ryd} for
a 20 \msun\ dwarf.
{\em CoStar} model A1 (solid line. Parameters as in 
Tab.~\protect\ref{ta_params} except $(\teff,\logg)$ = (35 kK, 4.0)), 
Kunze (dotted) :  $(\teff,\logg)$ = (35 kK, 4.0), and
Kurucz (dashed) : $(\teff,\logg)$ = (35 kK, 4.0)}
\label{fig_compare_20ryd}
\end{figure}
\begin{figure}[htb]
%\centerline{\psfig{figure=figs/compare_23ryd.eps,height=8.8cm}}
\centerline{\psfig{figure=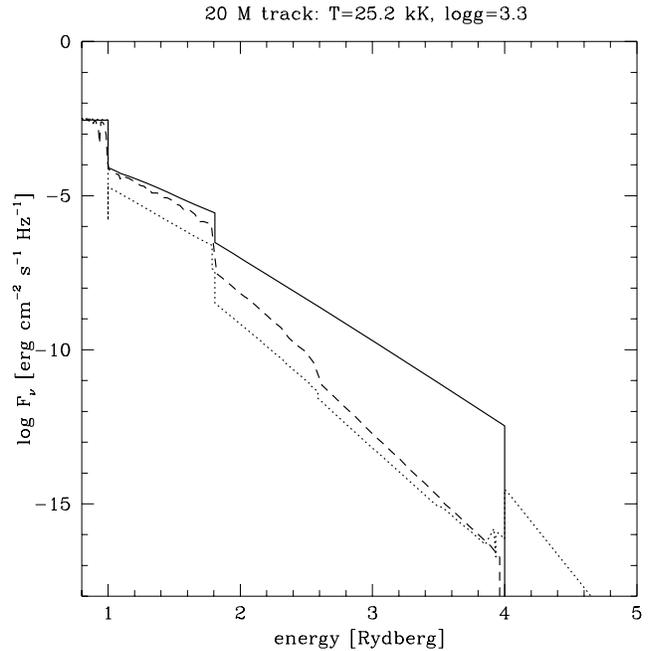,height=8.8cm}}
%\picplace{8.8cm}
\caption{Same as Fig.~\protect\ref{fig_compare_18ryd} for
a 20 \msun\ supergiant.
{\em CoStar} model A4 (solid line. Parameters as in 
Tab.~\protect\ref{ta_params})., 
Kunze (dotted) :  $(\teff,\logg)$ = (25 kK, 3.2), and
Kurucz (dashed) : $(\teff,\logg)$ = (25 kK, 3.3)}
\label{fig_compare_23ryd}
\end{figure}
\begin{figure}[htb]
%\centerline{\psfig{figure=figs/compare_23_temp.eps,height=8.8cm}}
\centerline{\psfig{figure=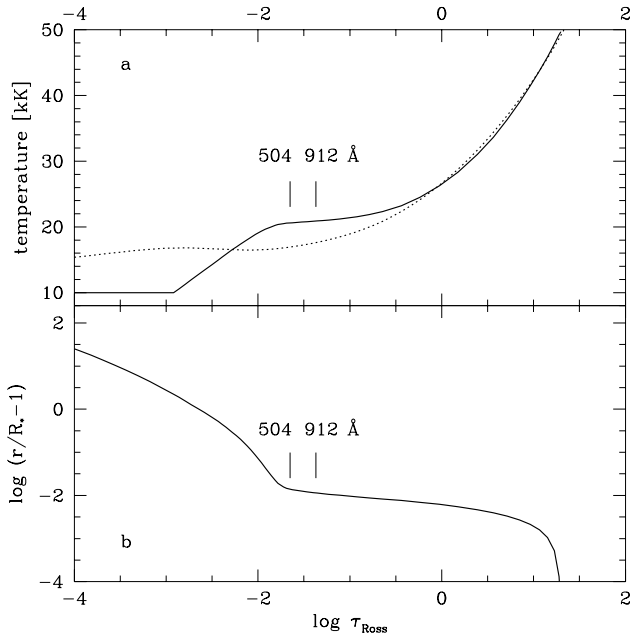,height=8.8cm}}
%\picplace{8.8cm}
\caption{{\bf a} Comparison of temperature structures from model A4 
(solid line) with a plane parallel \nlte\ model of Kunze (1994) for
$(\teff,\logg)$ = (25 kK, 3.2) (dotted).
Also indicated are the depths of the \he0\ and Lyman continuum forming 
layers ($\tau_\nu=2/3$) in the {\em CoStar} model A4.
The temperature differences are discussed in the text.
{\bf b} Measure of the spherical extension of model A4. Plotted is
$\log (r/\rstar-1)$ as a function of the Rosseland optical depth.
Note the location of the continuum forming layers in the 
photosphere--wind transition zone}
\label{fig_compare_23_temp}
\end{figure}

\subsubsection*{120 \msun\ track}

A comparison of models at the highest \teff\ available (approximately 
O3 V stars) is shown in Fig.~\ref{fig_compare_18ryd}. 
Plotted is model F2 from the 120 \msun\ track (solid line), 
the Kunze model for 
$(\teff,\logg)$ = (50 kK, 4.0) and Kurucz's model at 
$(\teff,\logg)$ = (50 kK, 5.0)\footnote{At high temperatures the 
results from Kurucz models do not depend sensitively on gravity 
and should thus allow for an appropriate comparison.}. 
While the Lyman continuum is essentially identical in all models, 
\nlte\ effects increase the flux in the \hei\ 
continuum in both \nlte\ models (e.g. Mihalas \& Auer 1970).
Given the very large temperature, H and He are fully ionized and electron
scattering becomes an important opacity source in the \hei\ continuum.
Consequently, wind effects
have little influence on the EUV flux up to the \heii\ continuum edge.
Close to the \heii\ edge, our model shows a small flux 
``excess'' with respect to the model of Kunze (1994). This is probably 
due to lower blanketing above the C~{\sc iii} and O~{\sc iii} edges in our 
{\em CoStar} models (see Sect.~\ref{s_improve}).
In the \heii\ continuum, the wind model shows the strong characteristic 
flux increase due wind effects (Cases {\em a-iii} and {\em c-ii}).

\subsubsection*{60 \msun\ track}

In Figures \ref{fig_compare_8ryd} and \ref{fig_compare_10ryd} we plot 
comparisons of dwarf and supergiant models from the 60 \msun\ track.

{\bf Dwarf model:}
Figure \ref{fig_compare_8ryd} shows the emergent flux 
for model D2 (solid line).
It is compared to a model with $(\teff,\logg)$ = (45 kK, 4.0) from Kunze
(dotted line) and a Kurucz model (dashed line) for (45 kK, 5.0).
In the Lyman continuum all models yield essentially the same results. 
In the \hei\ continuum, non-LTE effects increase the emergent flux.
The slope of the continuum of our {\em CoStar} model is slightly
flatter than that for the Kunze model (Case {\em a-ii}). This is
also found in recent calculations of Sellmaier \etal (1996).
 
In the \heii\ continuum the situation is a combination of Case {\em a-iii}
and {\em c-ii}.

{\bf Supergiant model:} 
The EUV flux distribution from the supergiant model (model D4 -- but recalculated 
for $(\teff,\logg)$ = (35 kK, 3.2)) is plotted in Fig.~\ref{fig_compare_10ryd} 
(solid line).
The comparison shows the corresponding plane parallel model from Kunze
for identical \teff\ and \logg\ (dotted line) and a Kurucz model 
(dashed) with $(\teff,\logg)$ = (35 kK, 4.0).
Given the strong stellar wind the ionizing flux is influenced by wind 
effects at essentially all wavelengths. As for the dwarf models, this 
leads to depopulated ground states (Case {\em a-iii}) but also causes the depth 
of continuum formation to be located in spherically extended layers
In this case the flattening of the spectrum in the \hei\ continuum
is mostly due to case ({\em b}).
The spectrum up to the \heii\ edge is flatter than plane parallel models. 
Due to stronger blanketing, the absolute flux in the Lyman 
continuum is lower than in the Kunze model.
Above the \heii\ edge the wind is optically thick up to large distances
(Case {\em b}), 
which explains the low flux with respect to the plane parallel \nlte\ model.

% 40 M track qualitatively same results ... but stronger effects !

\subsubsection*{20 \msun\ track}
\label{s_20track}

The EUV spectra along the 40 and 25 \msun\ tracks are 
qualitatively identical to the ones discussed above although the 
differences with respect to plane parallel \nlte\ models increase.
For the least massive objects modeled with {\em CoStar}, 
however, the situation may be slightly different due to their 
relatively low mass loss rates.

{\bf Dwarf model:} 
Figure \ref{fig_compare_20ryd} shows the dwarf 
model A1 (recalculated 
for $(\teff,\logg)$ = (35 kK, 4.0)) compared to Kunze and Kurucz models 
with identical parameters.
Due to stronger blanketing the Lyman continuum flux is slightly lower 
than the Kunze model.

Qualitatively the slope of the \hei\ continuum resembles the cases 
discussed before.
Figure \ref{fig_compare_20ryd}, however, shows that up to $\sim$ 
2.6--3 Rydberg the {\em CoStar} model predicts a flux lower than
that in  
both the \nlte\ and LTE plane parallel model.
The difference relative to the Kunze model is clear example of a
Case {\em a-ii} effect.
A similar result was also obtained for $\teff$ =35 kK by Sellmaier 
\etal (1996).
Due to the weak wind, the flux increase in the \heii\ continuum with 
respect to the plane parallel \nlte\ model is not strongly 
pronounced (Case {\em a-iii}).

{\bf Giant models:} 
Let us turn to the early B type giant models (Figure~\ref{fig_compare_23ryd}).
This star has a relatively weak wind, however, the {\em CoStar} model shows 
surprisingly large differences when compared to plane parallel models.
The cause for these large differences is not directly related to the
outflow (the \hi\ and \hei\ continua are formed in the photosphere), but
is a consequence of differences in temperature structure.

The temperature structures of model A4 and of
the Kunze model for $(\teff,\logg)$ = (25 kK, 3.2) are given in the upper panel of
Fig.~\ref{fig_compare_23_temp}. The lower panel shows
the spherical extension of model A4. 
Note that the structures agree well at optical depths \tauross $\ga$ 1. 
In the {\em CoStar} model the quasi-hydrostatic region reaches out to 
$\log \tauross \sim -1.7$. At this point the wind flow starts to accelerate,
causing the temperature to decrease rapidly.  
Given the weakness of the wind in the {\em CoStar} model, the intermediate 
region $-1.7 \la \tauross \la 1$ where geometric extension is negligible is
sufficiently large for the temperature 
to approach the value corresponding to that of the boundary temperature in 
a plane parallel grey atmosphere in radiative equilibrium, i.e. 
$T \rightarrow 0.841 \teff$. 
At optical depths $\log \tauross \la -1.7$, the spherical extension
makes itself known, leading to the rapid decline of the temperature.

As Fig.~\ref{fig_compare_23_temp} shows, the temperature in the plane 
parallel \nlte\ model is below the grey value, as expected, due to cooling
of the metals.
%lower in the intermediate range. 
The difference with {\em CoStar} reaches up to $\sim$ 3500 K close to the 
photosphere--wind transition zone, which is also the region where the 
\hei\ and the Lyman continuum are formed (see Fig.~\ref{fig_compare_23_temp}).
The higher temperature leads to a higher flux both in the Lyman and in 
the \hei\ continuum of the 
{\em CoStar} model with respect to the Kunze model 
(see Fig.~\ref{fig_compare_23ryd})\footnote{In both models the H and He groundstates
are overpopulated.}.
The Kurucz model yields results intermediate between both \nlte\ models.
As discussed by Kunze (1994) this is due to an overestimation of the temperature
gradient in the continuum forming layers in B star models of Kurucz 
(cf.~Philips \& Wright 1980).

In summary, we have seen that differences in the ionizing flux down to 228 \AA\
between {\em CoStar} models and plane parallel models become particularly
large for the lowest mass tracks presented here.
This is not what one would expect: 
Given the relatively low mass outflow of these objects, wind effects are
expected to become less and less important and the predictions should smoothly 
join those from \nlte\ plane parallel models\footnote{Non-LTE effects are 
still of importance and one therefore does not expect the results to
agree with the Kurucz models in this domain.}.

But this does not occur.
In fact the physical situation becomes more complex, as the temperature
structure start to play an important role.
Our predictions for BO dwarf and giant stars 
(roughly corresponding to models A) should thus be taken with caution. 
These uncertainties will be discussed in more detail in Sect.~\ref{s_improve}.

%%%%%%%%%%%%%%%%%%%%%%%%%%%%%%%%%%%%%%%%%%%%%%%%%%%%%%%%%%%%%%%%%%%%%%%%
\section{Revised ionizing fluxes of O and early B stars}
\label{s_revision}
In this section we present the integrated photon fluxes obtained from 
our models at solar and $1/5$ solar metallicity and derive new calibrations 
for O3 to B0 stars of population I.
We also compare our results with previous
studies that use different atmosphere calculations.

\subsection{Integrated photon fluxes}
In Table \ref{ta_qi_z020} we list the predicted number of photons 
emerging at wavelengths shorter than 912, 504, and 228 \ang\ respectively, 
referred to as $q_0$, $q_1$, and $q_2$. 
The values are given for both line blanketed models sets at Z=0.020 and
0.004.
The total ionizing luminosity $Q_i$ (in photons $\rm s^{-1}$) is
obtained by $Q_i=4 \pi (\rstar \rsun)^2 q_i$, where \rstar\ is given in Table 
\ref{ta_params}.

The $q_i$ values may be used to interpolate results for other stellar 
parameters and comparisons to predictions from plane parallel atmosphere 
models (cf.~below). It must, however, be kept in mind that the fundamental value
predicted by spherically extended models, such as the present ones, is the
total ionizing luminosity ($Q_i$), since the radius is one of the basic model 
parameters. 
Using interpolations of $q_i$ to obtain the number of ionizing photons for
stars with significantly differing radii may therefore yield 
unreliable results. The same comment also applies to the wind parameters.
Numerical care must be taken for interpolations since the model grid
is not uniformly spaced in its input parameters. While reliable $q_0$ and $q_1$ 
interpolations can be done with 2-D surface interpolation algorithms the same
is not true for $q_2$, which shows a strong discontinuity in the 
$\logg - \log\teff$ plane. For population synthesis applications the 
safest approach may be to assign the flux from some ``nearest neighbour'' 
model to the desired point and scale it to the correct bolometric luminosity, 
as mostly done.

{\footnotesize
\begin{table}[htb]
\caption{Ionizing photon fluxes in ${\rm cm^{-2} s^{-1}}$ from {\em CoStar} models
at Z=0.020 (columns 2-4) and Z=0.004 (columns 5-7).
For Z=0.020 the stellar parameters are given in Table \protect\ref{ta_params}. The low
Z models include abundance changes as well as the expected variations of the wind 
parameters.
Empty columns denote fluxes $< 10^3\, {\rm cm^{-2} s^{-1}}$)}
%
% NORMALISED values !
% *** verified, Feb 27, 1996
\centerline{
\begin{tabular}{rrrrrrrrrrr}
\\ \hline \\
      & \multicolumn{3}{c}{Z=0.020} & \multicolumn{3}{c}{Z=0.004}\\
model & $\log q_0$ & $\log q_1$ & $\log q_2$ & $\log q_0$ & $\log q_1$ & $\log q_2$\\
\\ \hline
\multicolumn{4}{l}{20 \msun\ track:} \\
A1  & 23.63 &	22.12	& 18.27	  & 23.73 &   22.33 & 18.75   \\ % 20 k0011 %* 
A2  & 23.35 &	21.59	&         & 23.39 &   21.69   \\ % 21 k0151 %*         
A3  & 22.92 & 	20.84	&	         & 22.96 &   20.95   \\ % 22 k0251 %*         
A4  & 21.89 &	19.10	&	         & 21.89 &   19.17   \\ % 23 k0351 %*         
\multicolumn{4}{l}{25 \msun\ track:}  \\        
B1  & 24.04 &	23.25 &	20.14	    & 24.10 &   23.33 & 20.03   \\ % 24 i0006 %* 
B2  & 23.85 &	22.85 &	19.67	    & 23.85 &   22.82 & 19.53   \\ % 25 i0101 %* 
B3  & 23.33 &	21.50 & 	    & 23.36 &   21.57   \\ % 26 i0211 %*         
B4  & 22.52 &	20.09 &		    & 22.52 &   20.08   \\ % 27 i0291 %*         
\multicolumn{4}{l}{40 \msun\ track:} \\	          
C1 & 24.47 &	23.97 &	21.56  	        & 24.50 &     24.00 & 20.49 \\ %* 1           
C2 & 24.33 &	23.75 & 21.27  	        & 24.35 &     23.80 & 20.69 \\ %* 2           
C3 & 24.05 &	23.31 &	20.13 	        & 24.08 &     23.40 & 20.13 \\ %* 3           
C4 & 23.00 &	20.84 &	8.69   	        & 23.23 &     21.25   \\ %* 4                 
C5 & 22.38 &	19.95 &	6.27  	        & 22.43 &     19.76   \\ %* 5                 
C6 & 20.90 &    14.02 &  	        & 20.75 &             \\ %* 6                 
\multicolumn{4}{l}{60 \msun\ track:} \\	                
D1 & 24.67 &	24.20 &	21.95 	        & 24.68  &   24.23 & 20.95 \\ %& 7            
D2 & 24.58 &	24.09 &	21.54   	        & 24.60  &   24.13 & 20.79  \\ %* 8           
D3 & 24.23 &	23.59 &	20.44	        & 24.25  &   23.64 & 20.64 \\ %* 9            
D4 & 23.64 &	22.48 &	12.42 	        & 23.64 &    22.50 & 11.49 \\ %* 10           
D5 & 22.68 &	20.18 &	7.09	        & 22.71 &    20.24 &  6.34 \\ %* 11           
D6 & 21.65 &	15.59 & 		& 21.65 &    18.16           \\ %* 12 q2=1.53 
D7 & 21.74 &    15.62 &		        & 21.82 &    18.34           \\ %* 13 q2=2.12 
\multicolumn{4}{l}{85 \msun\ track:} \\	          
E1 & 24.81 &	24.39 &	21.55 	        & 24.82 &    24.40 & 21.38 \\ %* 14           
E2 & 24.71 & 	24.25 & 21.49	        & 24.72 &    24.27 & 21.27 \\ %* 15           
E3 & 24.45 &	23.88 &	20.89 	        & 24.44 &    23.90 & 20.98 \\ %* 16           
\multicolumn{4}{l}{120 \msun\ track:} \\            
F1   & 24.92 &	24.50 &	21.79        & 24.91 &  24.51 & 21.57   \\ % 17 h0011 %* 
F2   & 24.80 &	24.35 &	21.74	     & 24.80 &  24.37 & 21.55   \\ % 18 h0071 %* 
F3   & 24.69 &	24.20 &	21.19	     & 24.69 &  24.22 & 21.53    \\ % 19 h0251 %*
\hline
\end{tabular}
}
\label{ta_qi_z020}
\end{table}
} % end footnotesize

\subsection{Metallicity effects}
\label{s_metals}
As mentioned in Sect.~\ref{s_z004}, a change in metallicity basically
affects our results in three different ways:
{\em 1)} Change of the total metal abundance,
{\em 2)} change of the relative hydrogen to helium abundances expected
from the chemical evolution, and 
{\em 3)} modified wind properties due to {\em 1)}.
In Table \ref{ta_qi_z020} we give the ionizing photon fluxes derived
from models at Z=0.004.
These values are to be compared to the ones for solar 
metallicity.

In most cases the ionizing flux in the H and \hei\
continua increases with lower metallicity, as expected. 
The difference between both sets is, however, relatively small.
Typically $q_0$ changes by less than $\sim$ 30 \%, 
%% ~ same order of magnitude changes as Kurucz !!
although larger variations are found for $q_1$.
The main influence in our models results from
effect {\em 3)}, mentioned above.
Effects {\em 1)} and {\em 2)} are of secondary importance.
At Z=0.004 both the decrease of the wind velocities and the
mass loss rate imply a diminishing importance of wind effects.
Consequently, cascading from upper levels (case {\em a-ii})
becomes less important. 
Thus the H groundstate is
less over-populated compared to the solar metallicity case, implying a 
slightly stronger flux in the Lyman continuum.
Similar reasoning applies to the \hei\
continuum for the temperature range covered in our models.

As discussed at length, the \heii\ ionizing flux is most sensitive
to wind effects. In general, a lowering of the metallicity implies a decrease 
of the wind density leading to less emission in the \heii\ continuum.
However, in some cases of stars with large wind densities (e.g.~models E3 and F3) 
the opposite is seen. This is similar to what is seen in hot Wolf-Rayet stars: 
The \heii\ continuum is formed at a relatively large radius, where 
the ground state population drives the ionization, making it proportional 
to the total density (e.g.~Schmutz \& Hamann 1986). In this situation
lowering the wind density (i.e.~lowering Z) reduces the recombination
rates, implying a stronger ionization. This explains the increase of
the \heii\ ionizing flux for the models with the highest wind densities
when lowering the metallicity\footnote{In model F3 the He enrichment
further augments this behaviour.}.
In summary, we see that the metallicity dependence of the ionizing 
fluxes predicted by our models is relatively small and essentially due
to the expected changes of the wind properties with Z 
(effect {\em 3)}).
However, given the likely underestimate of blanketing in our models (see 
Sect.~\ref{s_improve}) we expect the  
dependence on metal abundances to be somewhat more important than found here.

%%% BELONGS to next subsection 
{\footnotesize
\begin{table*}
\caption{Parameters for OB-type stars: Ionizing photon fluxes per unit surface area
($q_0$ and $q_1$ in column 4 and 5) derived from solar metallicity {\em CoStar} 
models, using the \teff--\logg\ calibration of Vacca \etal (1996) given in
columns 2 and 3. 
Adopting the radii from Vacca \etal\ (column 4) one obtains the absolute
photon fluxes $Q_0$ and $Q_1$ given in column 5 and 6}
\centerline{
\begin{tabular}{lrrrrrrrrrr}
\\ \hline \\
Sp.~Type & \teff & $\log g_{\rm evol}$ & $\log q_0$ & $\log q_1$ & $R$ & $\log Q_0$ & $\log Q_1$ \\ 
	 & [K]	 & [cgs]  & [${\rm cm^{-2} s^{-1}}$] & [${\rm cm^{-2} s^{-1}}$] & [\rsun] 
	 & [$\rm s^{-1}$] & [$\rm s^{-1}$] \\ \hline \\
\multicolumn{4}{l}{Luminosity class V:} \\
%%% New calibration fro Vacca et al. (accepted version of paper -has slightly
% different Teff, logg values...
%  Data as from october 4, 1995
% from unit.14 (output of test_ionis)
% NAG rnw:   0.55315518915049   0.55315518915049   0.55315518915049
% Ionizing fluxes for OB stars (CoStar)
% --> parameters from Table 5-7 using logg_evol !
% 
% *** verified, Feb 27, 1996
%
%  Sp        Teff     	logg        q_0       q_1     R	     Q_0     Q_1
O3   V     &  51230. &   4.149  &   24.82 &   24.39 & 13.2 &   49.85 &   49.42 \\
O4   V     &  48670. &   4.106  &   24.71 &   24.27 & 12.3 &   49.68 &   49.23 \\
O4.5 V     &  47400. &   4.093  &   24.65 &   24.20 & 11.8 &   49.58 &   49.12 \\
O5   V     &  46120. &   4.081  &   24.58 &   24.11 & 11.4 &   49.48 &   49.01 \\
O5.5 V     &  44840. &   4.060  &   24.51 &   24.00 & 11.0 &   49.38 &   48.86 \\
O6   V     &  43560. &   4.042  &   24.44 &   23.91 & 10.7 &   49.28 &   48.75 \\
O6.5 V     &  42280. &   4.030  &   24.36 &   23.81 & 10.3 &   49.17 &   48.62 \\
O7   V     &  41010. &   4.021  &   24.27 &   23.65 & 10.0 &   49.05 &   48.44 \\
O7.5 V     &  39730. &   4.006  &   24.18 &   23.50 & 9.6  &   48.93 &   48.25 \\
O8   V     &  38450. &   3.989  &   24.08 &   23.33 & 9.3  &   48.80 &   48.05 \\
O8.5 V     &  37170. &   3.974  &   23.94 &   23.05 & 9.0  &   48.64 &   47.74 \\
O9   V     &  35900. &   3.959  &   23.79 &   22.70 & 8.8  &   48.46 &   47.37 \\
O9.5 V     &  34620. &   3.947  &   23.60 &   22.27 & 8.5  &   48.25 &   46.92 \\
B0   V     &  33340. &   3.932  &   23.40 &   21.79 & 8.3  &   48.02 &   46.41 \\
B0.5 V     &  32060. &   3.914  &   23.18 &   21.27 & 8.0  &   47.77 &   45.86 \\
\\
\multicolumn{4}{l}{Luminosity class III:} \\
O3   III   &  50960. &   4.084   &   24.81 &   24.37 & 15.3 &   49.97 &   49.52 \\
O4   III   &  48180. &   4.005   &   24.70 &   24.24 & 15.1 &   49.84 &   49.38 \\
O4.5 III   &  46800. &   3.971   &   24.64 &   24.18 & 15.0 &   49.78 &   49.32 \\
O5   III   &  45410. &   3.931   &   24.58 &   24.11 & 15.0 &   49.71 &   49.25 \\
O5.5 III   &  44020. &   3.891   &   24.51 &   24.03 & 14.9 &   49.64 &   49.16 \\
O6   III   &  42640. &   3.855   &   24.43 &   23.92 & 14.8 &   49.56 &   49.05 \\
O6.5 III   &  41250. &   3.820   &   24.34 &   23.78 & 14.8 &   49.47 &   48.91 \\
O7   III   &  39860. &   3.782   &   24.24 &   23.63 & 14.7 &   49.36 &   48.75 \\
O7.5 III   &  38480. &   3.742   &   24.12 &   23.41 & 14.7 &   49.24 &   48.53 \\
O8   III   &  37090. &   3.700   &   23.97 &   23.03 & 14.7 &   49.09 &   48.14 \\
O8.5 III   &  35700. &   3.660   &   23.82 &   22.68 & 14.7 &   48.94 &   47.80 \\
O9   III   &  34320. &   3.621   &   23.64 &   22.28 & 14.7 &   48.76 &   47.40 \\
O9.5 III   &  32930. &   3.582   &   23.44 &   21.83 & 14.7 &   48.56 &   46.95 \\
B0   III   &  31540. &   3.542   &   23.21 &   21.35 & 14.7 &   48.33 &   46.47 \\
B0.5 III   &  30160. &   3.500   &   22.98 &   20.90 & 14.8 &   48.11 &   46.03 \\
\\
\multicolumn{4}{l}{Luminosity class I:} \\
O3   I     &  50680. &   4.013 &   24.81 &   24.34  & 17.8 &   50.09 &   49.63 \\
O4   I     &  47690. &   3.928 &   24.69 &   24.24  & 18.6 &   50.02 &   49.56 \\
O4.5 I     &  46200. &   3.866 &   24.63 &   24.19  & 19.1 &   49.98 &   49.53 \\
O5   I     &  44700. &   3.800 &   24.57 &   24.10  & 19.6 &   49.94 &   49.47 \\
O5.5 I     &  43210. &   3.740 &   24.49 &   23.96  & 20.1 &   49.88 &   49.35 \\
O6   I     &  41710. &   3.690 &   24.40 &   23.83  & 20.6 &   49.81 &   49.24 \\
O6.5 I     &  40210. &   3.636 &   24.30 &   23.69  & 21.2 &   49.73 &   49.12 \\
O7   I     &  38720. &   3.577 &   24.18 &   23.45  & 21.8 &   49.64 &   48.91 \\
O7.5 I     &  37220. &   3.516 &   24.05 &   23.17  & 22.4 &   49.53 &   48.65 \\
O8   I     &  35730. &   3.456 &   23.91 &   22.86  & 23.1 &   49.42 &   48.37 \\
O8.5 I     &  34230. &   3.395 &   23.75 &   22.52  & 23.8 &   49.29 &   48.05 \\
O9   I     &  32740. &   3.333 &   23.55 &   22.11  & 24.6 &   49.12 &   47.67 \\
O9.5 I     &  31240. &   3.269 &   23.31 &   21.61  & 25.4 &   48.90 &   47.21 \\
\hline
\end{tabular}
}
\label{ta_vacca}
\end{table*}
} % end footnotesize
%\clearpage

\subsection{A new ionizing flux calibration for O3--B0 stars}
Recently Vacca \etal (1996) derived new calibrations of stellar parameters
for O3 to B0.5 stars. These are based on results from detailed modeling of
the observed absorption line spectra of stars with well-defined spectral
classifications. Using these calibrations Vacca \etal (1996)
calculate ionizing photon fluxes in the H and \he0\ continua based
on Kurucz (1991) LTE model atmospheres\footnote{For clarification a brief 
comment on the procedure of Vacca \etal (1996) for deriving Kurucz ionizing 
fluxes appears useful: A shown in Fig.~\ref{fig_hr_logg_models} the 
Kurucz (1991) models only cover the domain of the 20 and 25 \msun\ track.
It is important to note that for the rest of the domain of interest, which 
corresponds to all O3 to O9 stars, extrapolations combined
with blackbody spectra are used to extend the Kurucz grid to the 
desired lower gravities. 
}.
In view of the
important improvements included in our {\em CoStar} models, it
is of particular interest to provide a recalibration of the photon fluxes
taking into account \nlte and wind effects and \lb.

Table \ref{ta_vacca} shows our new calculations of the H and \hei\ 
ionizing fluxes for O3 to B0.5 stars of luminosity classes V, III, and I.
Given are the photon fluxes per unit surface $q_0$ and $q_1$. 
We also provide absolute photon fluxes $Q_0$ and $Q_1$, adopting the 
radii from the recent calibration of Vacca \etal (1996) 
in Table \ref{ta_vacca}.
The  photon fluxes have been derived from our solar metallicity models 
(Table \ref{ta_qi_z020}) using
a two-dimensional surface interpolation routine based on a modified 
Shepard method (NAG {\tt e01sef} routine). For the reasons mentioned
above we have refrained from interpolating values for the \heii\ ionizing
fluxes.
The $\log q_i$ values have been interpolated 
in the $\logg-\log\teff$ plane. 
For the gravity calibration we prefer to use the ``evolutionary gravity''
$g_{\rm evol}$ given by Vacca \etal, since these values are consistent
with the gravity definition in our {\em CoStar} models. 
Furthermore the distinction with their ``spectroscopic gravity'' is not
of great concern for the comparison with the ionizing fluxes 
since their values are only weakly dependent on 
the adopted gravity (see below).

\begin{figure*}[htb]
%\centerline{\psfig{figure=figs/compare_q0q1.eps,height=8.8cm}
%\psfig{figure=figs/costar_panagia2.eps,height=8.8cm}}
\centerline{\psfig{figure=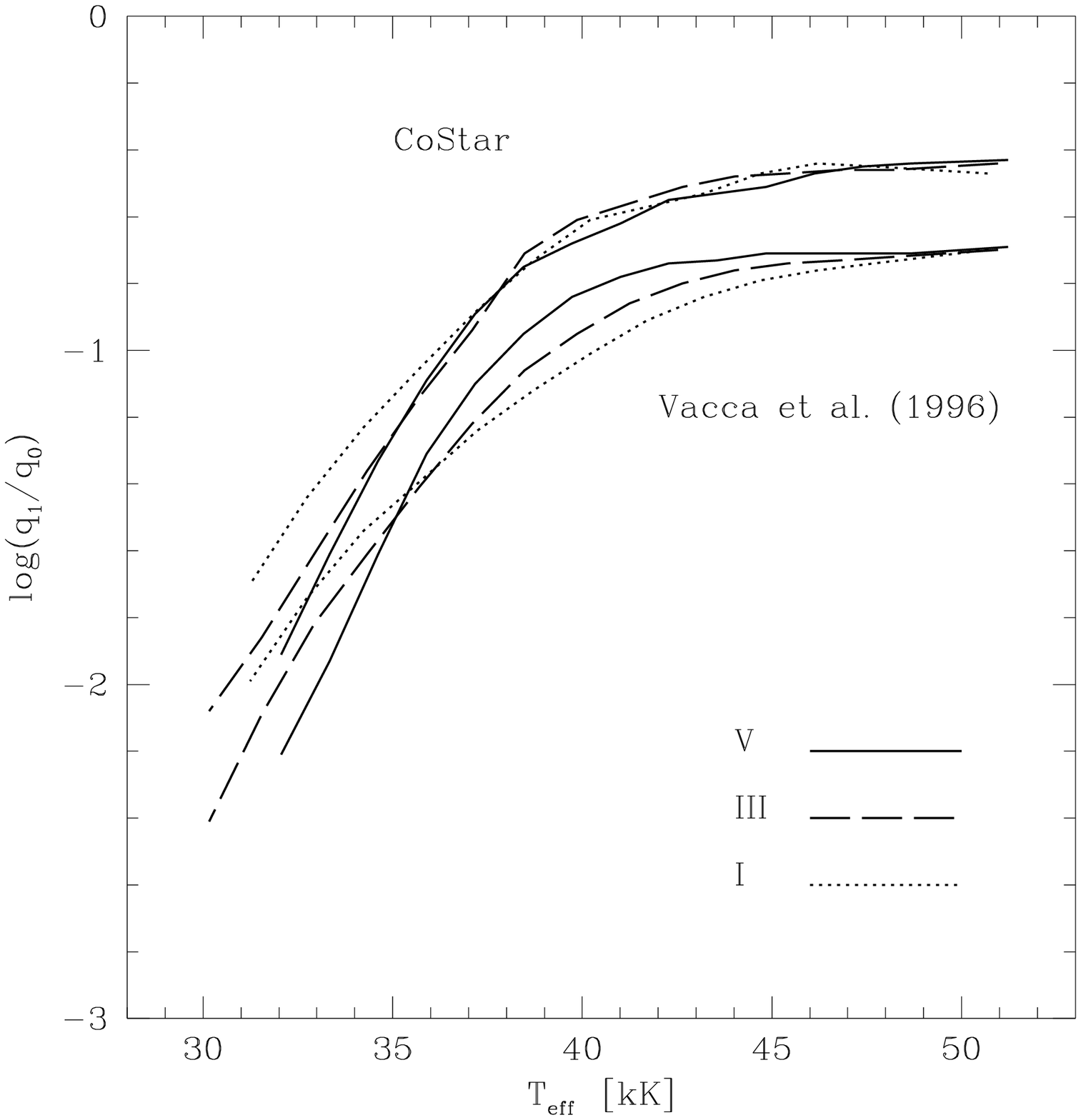,height=8.8cm}
\psfig{figure=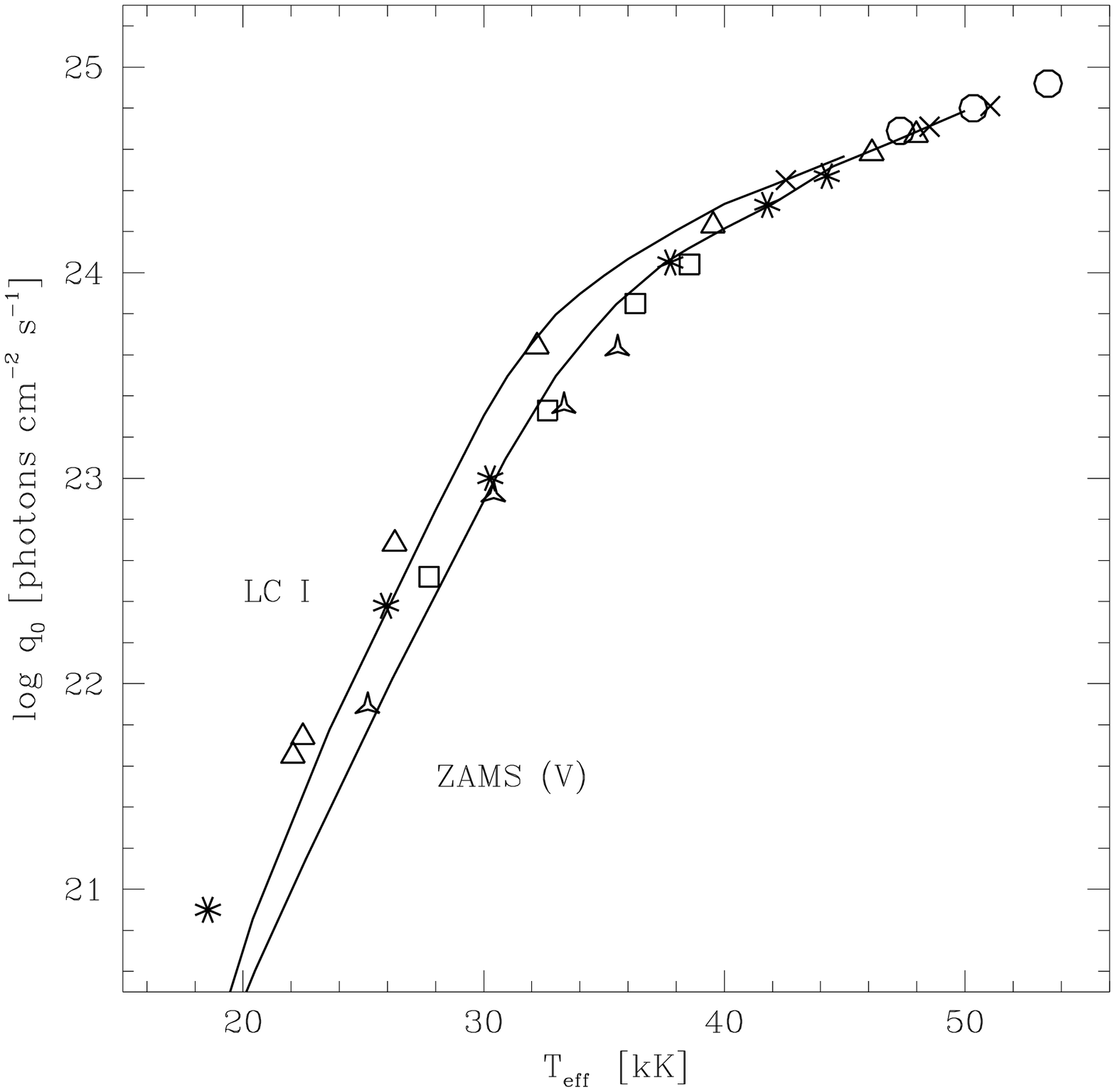,height=8.8cm}}
%\picplace{8.8cm}
\caption{{\bf Left panel:} 
Logarithm of the hardness ratio $\log\left(q_1/q_0\right)$
as a function of effective temperature for dwarfs, giants and 
supergiants. Shown are the values from our {\em CoStar} models
(see Table \protect\ref{ta_vacca}) and those derived by Vacca \etal
(1996) using Kurucz models. 
%See text for discussion.
{\bf Right panel:}
Logarithm of the number of Lyman continuum ionizing photons 
versus effective temperature. 
The predictions from our models are plotted using the same symbols as
in Fig.~\protect\ref{fig_hr_logg_models}.
The solid lines shows the relations from Panagia (1973) for his ZAMS 
and luminosity class I}
\label{fig_compare_q0q1}
\end{figure*}

\subsection{Comparison with previous calibrations}
We compare our results from Table \ref{ta_vacca} with 
those derived by Vacca \etal (1996) using the Kurucz (1991) models.
A brief comparison with earlier work of Panagia (1973) will also be given.

For all O3 to B0.5-type stars, the total number of {\em Lyman continuum photons} 
$q_0$ is somewhat {\em lower} than predicted by the Kurucz models 
The difference
increases from 0.03 to 0.07 dex ($\sim$ 7 \% to 15 \%) between O3 and O7 respectively. 
For later types this behaviour increases strongly and reaches 0.14 (0.26) 
dex  			%% $\sim$ 40 (80) \% 
for  B0 dwarfs (O9.5 supergiants).
More important is the comparison with the \hei\ photon flux, which
is strongly affect by \nlte\ and wind effects (see above).
For the dwarf sequence in Table \ref{ta_vacca}, {\em $q_1$} is 
{\em larger} by $\sim$ 25 to 75 \% with respect to the values 
presented by Vacca et al. For supergiants $q_1$ is typically 
increased by a factor of 2.
It has already been noted
by Vacca (1994) and by Vacca \etal (1996)
that the $q_1$ values based on Kurucz models should be systematically
underestimated because of the assumption of LTE.

As mentioned in Sect.~\ref{s_20track}, our results for
the latest spectral types do neither smoothly tend towards the Kurucz 
values, nor do they tend to the results of the plane parallel
\nlte\ models of Kunze (1994).
Model calculations for later types will be necessary to locate more
precisely where \nlte\ and wind effects become negligible
(but see Sect.~\ref{s_improve}).
Since for most applications involving \hii\ regions and alike systems the 
ionizing spectrum will be determined by the most massive 
stars present, the presence of the ``discontinuity'' will not
be of importance.

In Figure \ref{fig_compare_q0q1} (left panel) we plot the hardness 
ratio $q_1/q_0$ from the {\em CoStar} values of Table \ref{ta_vacca} 
as a function of \teff. Also shown is the hardness ratio derived by Vacca 
\etal.
As pointed out by these authors (cf.~also Vacca 1994), it must  
be kept in mind that these values should be higher by typically
a factor of 2 if one accounted for \nlte\ effects in plane parallel models.
Figure  \ref{fig_compare_q0q1} clearly shows the overall 
increase of the hardness of the ionizing spectrum at a
given temperature. The strongest hardening is found
for supergiants due to the increasing importance of wind effects.
Interestingly, for stars with \teff $\ga$ 36000 K we find that the 
hardness ratio is essentially independent of luminosity class while 
the values of Vacca \etal show a spread of up to $\sim$ 0.2 dex.
To produce a hardness ratio of $\log\left(q_1/q_0\right) < -0.7$ 
with our new models the temperature of the exciting star can be $\sim$ 
1000 to 11000 K smaller than if Kurucz models were used. 
Therefore the general tendency is that lower effective temperatures 
would be derived from observed nebular properties if one uses 
atmosphere models, which account for \nlte, \lb\ and stellar winds.

%%%%% figures belong to next section ! 
\begin{figure}[htb]
%\centerline{\psfig{figure=figs/eps_cma2.eps,height=8.8cm}}
\centerline{\psfig{figure=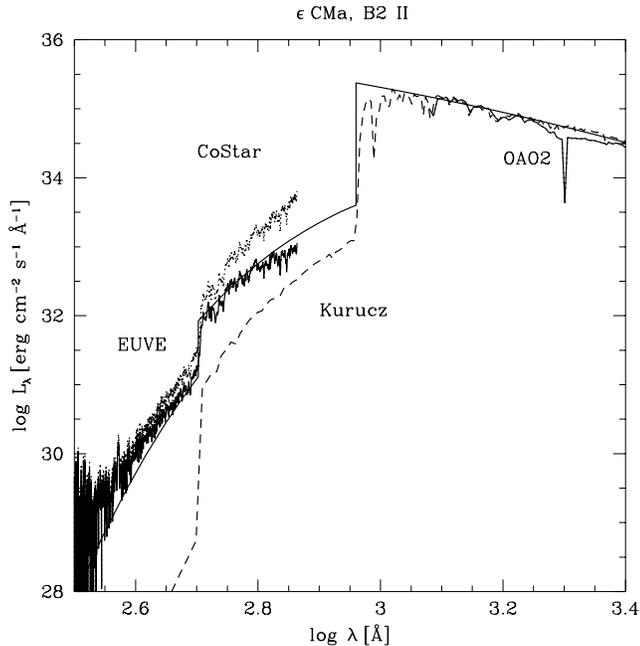,height=8.8cm}}
%\picplace{8.8cm}
\caption{Far UV and EUV flux of $\epsilon$ CMa. EUVE observations from 
Cassinelli \etal (1995) corrected  for an attenuation by 
$N_{\rm H \, I}=1.\, 10^{18}$ (dotted line up to $\sim$ 700 \ang)
and $N_{\rm H \, I}=5.\, 10^{17} \; {\rm cm^{-2}}$ (Gry \etal 1995; solid
line)
Model comparisons: {\em CoStar} model (solid line, Parameters from Table
\protect\ref{ta_cma}), Kurucz model (dashed) with $(\teff,\logg)$ = (21 kK, 3.0).
All fluxes are scaled (consistently with \rstar) to the distance of 188 pc.
Discussion in text}
\label{fig_eps_cma1}
\end{figure}
% BELONGS to next subsection !
\begin{figure}[htb]
%\centerline{\psfig{figure=figs/beta_cma1.eps,height=8.8cm}}
\centerline{\psfig{figure=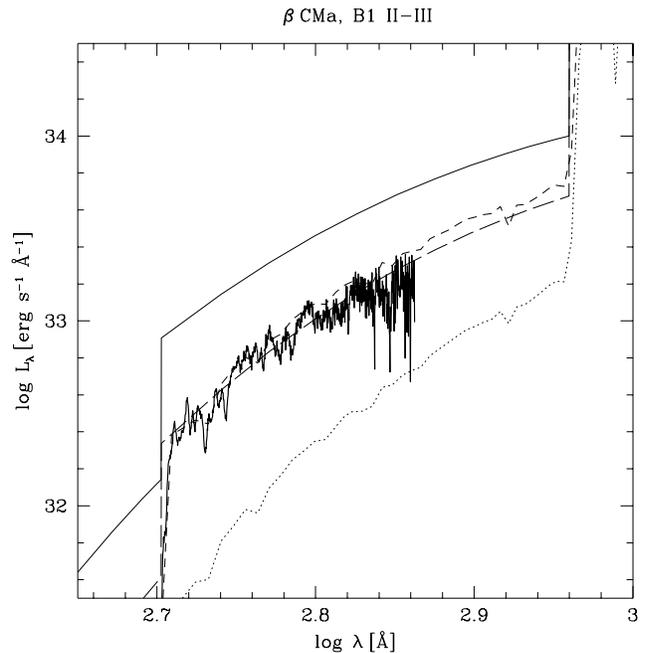,height=8.8cm}}
%\picplace{8.8cm}
\caption{EUV flux of $\beta$ CMa. EUVE observations from 
Cassinelli \etal (1996) corrected  for an attenuation by 
$N_{\rm H \, I}=2.\, 10^{18} \; {\rm cm^{-2}}$ (solid line up to $\sim$ 700 \ang).
The following models are shown for comparisons: {\em CoStar} models (Parameters 
from Table \protect\ref{ta_cma}. Solid line: hot model ; long-dashed: cooler model), 
Kurucz models with $(\teff,\logg)$ = (25 kK, 3.5) (dashed) and
(23 kK, 3.5) (dotted).
All fluxes are scaled (consistently with \rstar) to the distance of 206 pc.
Discussion in text}
\label{fig_beta_cma1}
\end{figure}

Finally it is interesting to make a brief comparison with the widely used
results of Panagia (1973). For further comparisons see Vacca et al.
Figure \ref{fig_compare_q0q1} (right panel) shows the comparison of
the number of Lyman continuum photons between Panagia and our
{\em CoStar} models. The behaviour of the models from the 40, 60, and
85 \msun track has already been discussed in Paper II.
The figure clearly illustrates the smaller
{\em CoStar} Lyman continuum flux for the dwarfs at $\teff \la 40$ kK 
which was also apparent in the above comparison with Vacca et al.
The same also holds for the evolved stars from the 20 and 25 \msun
tracks.
The models with values of $q_0$ larger than the supergiants
of Panagia are the extreme supergiant models from the 40 and
60 \msun track.

%%%%%%%%%%%%%%%%%%%%%%%%%%%%%%%%%%%%%%%%%%%%%%%%%%%%%%%%%%%%%%%%%%%%%%%%
\section{Discussion}
\label{s_improve}

We have shown that our models provide an important
improvement describing the ionizing fluxes of hot stars. However, it is 
also important to discuss possible shortcomings in our method, which may 
affect our results.
As discussed in Sects.~\ref{s_input} and \ref{s_costar} the uncertainties 
in our models are expected to become increasingly important for B stars, 
which have weak winds. 
This can be easily understood and it is best illustrated by considering
the following brief exploratory study of two unique cases of B2 giants, 
which have recently been observed shortward of the Lyman limit with EUVE.

%%%%%% beginning of new EUVE session:
\subsection{Comparisons with EUVE observations}
\label{s_euve}

The first EUV spectra from early type stars 
have been observed by Hoare \etal (1993).
Subsequent EUVE observations of $\epsilon$ CMa (B2II) and $\beta$ CMa 
(B1 II-III) by Cassinelli \etal (1995, 1996), have provided spectra allowing 
detailed comparisons with model atmospheres. 
As mentioned before, these observations have revealed important discrepancies with
predictions from plane parallel model atmospheres.
Najarro \etal (1996) have pointed out that these shortcomings could,
at least partly, be explained by models which account for stellar winds.
Their invoked mechanism is that of Case {\em a-iii}~working in the \hei\
continuum. This is the same mechanism that is relevant in the \heii\
continua of the O-type star models discussed here. 
It is therefore interesting to see how well our predictions agree, 
notwithstanding the fact that the stellar parameters of $\epsilon$ and $\beta$ 
CMa lie somewhat outside of the temperature domain covered in this work.

\subsubsection{$\epsilon$ CMa}

The adopted stellar parameters for $\epsilon$ and $\beta$ CMa
are given in Table \ref{ta_cma}.
Temperature, gravity and radius are from Cassinelli \etal (1995). 
We assume solar abundances.
The wind parameters of $\epsilon$ CMa are somewhat uncertain.
The adopted mass loss rate for $\epsilon$ CMa is compatible with the studies 
of Drew \etal (1994), Cassinelli \etal (1995) and Najarro \etal (1996). 
The same \vinf\ and $\beta$ as assumed by Najarro \etal are chosen to 
allow for a comparison with their work.

\begin{table}[htb]
\caption{Stellar parameters for $\epsilon$ and $\beta$ Canis Majoris}
\centerline{
\begin{tabular}{lrrrrrrrrr}
\hline \\
 & \multicolumn{1}{c}{$\epsilon$ CMa} & \multicolumn{1}{c}{$\beta$ CMa} \\
\hline \\
\teff\ [kK] &	21.0  	&  (24.8, 23.25) \\
\logg\ [cgs] &	3.2  	&  (3.7, 3.4) \\
%\distance & 	188 pc \\
\rstar\ [\rsun ] &	16.2  & 11.54 \\
$n_{\rm H}$ &	0.9  & 0.9 \\
\mdot\ [\masl ]	&	$1. \, 10^{-8}$ & $3.5 \, 10^{-8}$ \\
\vinf\ [\kms ]	&	800.  & 1220. \\
$\beta$	&	1.	& 1.
\\ \hline
\end{tabular}
}
\label{ta_cma}
\end{table}

The spectral energy distribution predicted by {\em CoStar} and the one
observed are given in Fig.~\ref{fig_eps_cma1}.
The result from a Kurucz model (dashed line) for 
$(\teff,\logg)$ = (21 kK, 3.0) plotted for comparison.
The dotted line shows the observed
EUVE flux corrected for a
hydrogen column density $N_{\rm H \, I}=1.\, 10^{18} \; {\rm cm^{-2}}$ 
(cf.~Cassinelli \etal 1995). The solid line shows the observation corrected
using the recently derived value of
$N_{\rm H \, I}=5.\, 10^{17}$ (Gry \etal 1995).
The comparison with the plane parallel LTE model illustrates the striking 
underestimate
of the continuum flux in both the Lyman and the \hei\ continuum. 
As pointed out by Cassinelli et al.,
the finding of an observed ``EUV excess'' holds for any plane parallel model,
i.e. it is
independent of the assumption of LTE or \nlte. \footnote{Non-LTE effects 
even decrease the EUV flux for the parameters of interest.}
As the figure shows, our model atmosphere yields a stronger EUV
emission which significantly improves the comparison with the observations.

{\em Does this result solve the ``EUV excess'' problem in $\epsilon$~CMa? }
The answer is: {\em no}. To explain this conclusion we need to identify the 
reason(s) for the EUV increase in our model relative to that of the plane 
parallel model. There are two basic reasons:

The first and major reason is simply a larger temperature
in the continuum forming layers of our {\em CoStar} model.
Indeed, in our model the temperature at the depth of continuum formation
at $\lambda \sim$ 600 \ang\ is $T \simeq$ 17 kK, which is larger than the 
value in the 
Kurucz model (T $\simeq$ 14 kK, Cassinelli \etal 1995). Although the ground states 
of H and He are slightly overpopulated in this region, the net result is 
nevertheless a stronger EUV emission.
In fact, the structure of our model qualitatively resembles
the structure of model A4 plotted in Fig.~\ref{fig_compare_23_temp}.
The considered continuum forming layers are located at small optical depths 
($\log\tauross \la -2$), where the temperature drop due to the spherical 
extension has not yet set in and the temperature is thus close to the 
boundary value of
the plane parallel grey atmosphere. This explains
the relatively high temperature. As for model A4, discussed
in Sect.~\ref{s_20track}, we therefore conclude that the EUV emission is 
strongly dependent on the temperature structure 
in the transition zone between photosphere and wind.

The second reason regards the treatment of line broadening and its
effect on the amount of line blanketing.
In the present calculations for $\epsilon$ and $\beta$ CMa, 
line blanketing turns out to be essentially ineffective.
This unrealistic behaviour is due to the neglect of line broadening 
(cf.~below). 
Introducing a small ``turbulent'' line broadening using the simple
method of Schaerer \& Schmutz (1992) results in a significant
decrease of the EUV flux.
This further points out the need for an improved treatment of \lb\ 
for later type stars with weak stellar winds.

Najarro \etal (1996) have pointed out the importance of the stellar wind on 
the ground state populations of H and He and hence on the emergent EUV flux
of early B giants. 
To compare our models with their results we calculated a series of models
where we varied the mass loss rate from $10^{-9}$ to $10^{-6}$ \masl, keeping
all other parameters as in Table \ref{ta_cma}.
We do not confirm their strong dependence of the model fluxes on mass loss.
In particular we obtain a much weaker dependence of the Lyman jump, the \hei\
jump and the number of ionizing photons on mass loss. 
This finding is confirmed by comoving frame calculations 
(Schmutz 1995, private communication).
Our results thus indicate that the strong mass loss dependence of the 
EUV flux found by Najarro \etal\ must be partly due to variations of their
temperature structure with \mdot. 
The amplitudes of these variations seem significantly larger than those in
our calculations.
This may originate from their simplified energy equation (they assume radiative 
equilibrium accounting only for H and He), which is used to derive 
their temperature structure.
This finding stresses the importance of deriving both reliable temperature 
structures and accurate \nlte\ populations including a detailed treatment of 
\lb\ and stellar winds to obtain reliable ionizing fluxes for B stars.

\subsubsection{$\beta$ CMa}

The adopted stellar parameters for $\beta$~CMa are given in Table \ref{ta_cma}.
Given the ambiguity of these parameters (Cassinelli \etal 1996), 
we present two {\em CoStar} models corresponding to the limiting cases discussed
by Cassinelli et al.
For the wind parameters we also follow these authors and adopt their theoretical 
estimates based on the modified CAK theory. 
As for $\epsilon$ CMa we assume solar abundances.

The predicted spectra and the {\em EUVE} observations are compared in
Fig.~\ref{fig_beta_cma1}.
The solid and long-dashed lines show the results from the hotter and 
cooler {\em CoStar} model respectively.  Kurucz model predictions for
comparable parameters (see figure caption) are shown as short-dashed and 
dotted lines.
Compared to the Kurucz models, our calculations again show a stronger
EUV flux for a given \teff. Similar to the $\epsilon$~CMa model discussed above,
this is mostly due to a temperature difference in the continuum 
forming layer as the ground state of H is overpopulated. 

The {\em CoStar} model with the lower \teff\ (23250 K) fits the {\em EUVE} 
observations best. 
Interestingly, this effective temperature was favoured by Cassinelli 
\etal (1996) based on a comparison of the UV to near-IR flux distribution.
In this case, the EUV emission predicted by the Kurucz model
is considerably too weak. Our results thus show that even for the
low \teff\ value and despite \nlte\ effects, which overpopulate the ground 
state, the observed EUV flux can be reproduced with the temperature
structure from our models. As for $\epsilon$~CMa, the key questions are again: 
how realistic is this structure, and more fundamentally, what 
are the physical processes that establish such a temperature structure?

%%%%%% END of new EUVE session:

\subsection{Current approximations and future improvements}
The exploratory results in the previous section clearly show that for stars 
of spectral types later than approximately B0 (not covered by our grid)
reliable predictions of ionizing fluxes are not yet possible. 
Therefore we have limited our model set to O3-B0 stars.
We shall now briefly discuss the most important model assumptions
and their importance for the set of calculations presented in Table
\ref{ta_params}.
The assumptions listed in Sect.~\ref{s_input} will be addressed in the 
following.

The Sobolev approximation, which is made for the line transfer
should yield sufficiently precise results for the parameters
of our model set (see also de Koter \etal 1993).
The weakest point in our Monte-Carlo treatment of line blanketing is
most likely the neglect of line broadening, yielding only a poor treatment
of photospheric lines (see Paper II). Line blanketing in the low velocity part 
of the atmosphere is therefore underestimated in the present models.
Although we have presently no possibility to quantify the importance
of photospheric blanketing, we expect that our results should be 
fairly reliable for the O stars for the following reasons:
{\em 1)} photospheric lines are both weaker and less numerous than in later 
types, and
{\em 2)} given their strong outflow, wind effects play a dominant role in 
establishing the equilibrium population.

A characteristic feature of the line blanketed \nlte\ models of Kunze
(1992, 1994) is the appearance of relatively strong absorption edges
in the \hei\ continuum which are mostly due to CNO, but also due to Ne and Ar
(see e.g.~Figs.~\ref{fig_compare_10ryd}, \ref{fig_compare_20ryd}).
These edges are not treated in our calculations.
The Kunze models, on the other hand, do not include lines of
iron peak elements, which are treated in our calculations, and which cause
an important fraction of the metal line blanketing.
Recently Sellmaier \etal (1996) presented calculations which
include line blocking and \nlte\ effects in a stellar wind model.
We note that their exploratory results also lack of pronounced 
metal ionization edges in the EUV, which seems to confirm our calculations.
A more detailed analysis of this question will be possible with
the inclusion of additional metals in the full \nlte\ calculations
(see e.g.~de Koter \etal 1994, 1996a).

A potential source of uncertainty for high \teff\ models could come
from the coherent treatment of electron scattering, which might modify
the He ionization, as pointed out by Rybicki \& Hummer (1994).
This effect remains to be included in future models.

As discussed in Paper II, the high energy part of the spectrum may
be affected by the emission of X-rays which are usually
attributed to shocks in the stellar winds.
We did not include such shocks in our calculations. This may cause
us to underestimate the flux in the \hep\ continuum.
At longer wavelengths, our results should hardly be affected
by X-rays as shown by the work of MacFarlane \etal (1994).
They find that X-rays 
cause only a small perturbation of the wind structure in the O-type stars
discussed in this paper.
For stars of later spectral types the situation is different.
For these stars, which have relatively weak winds, they
do find that X-rays
may drastically alter the wind ionization and possibly also 
contribute to heating in the photosphere.
For the understanding of later types than those included in our sets, 
and particularly for an explanation of the ``EUV excess'' of $\epsilon$ 
and $\beta$ CMa (cf.~Sect.~\ref{s_euve}), the inclusion of X-rays will 
probably be of great concern.

%%%%%%%%%%%%%%%%%%%%%%%%%%%%%%%%%%%%%%%%%%%%%%%%%%%%%%%%%%%%%%%%%%%%%%%%
\section{Summary and conclusions}
The present work provides an extensive set of predictions regarding
the spectral energy distribution of massive stars ($M_{\rm i}=$
20 to 120 \msun)
derived from our combined stellar structure and atmosphere models.
Our set covers the entire main sequence evolution and approximately
corresponds to O3--B0 stars of all luminosity classes.
This represents the first set of predictions for O and 
early-B stars, which are based on the most recent atmosphere models 
accounting for \nlte\ effects, line blanketing, and stellar winds.
Especially the treatment of the stellar wind is found to be of
great importance for predicting reliable ionizing fluxes
of hot stars. Our calculations should provide, for the first
time, a reliable description of the spectral energy distribution
in the \hei\ and \heii\ continuum, where \nlte, \lb, and wind effects
are crucial.

We have discussed the 
importance of wind effects and \lb\ in
O3 to B0 stars in relation to the predicted ionizing fluxes.
As already partly found in previous investigations 
(Gabler \etal 1989, 1992; Schaerer 1995, Najarro \etal 1996, 
Schaerer \etal 1996a, b) these effects have a profound importance
on shaping the EUV flux of hot MS stars.
The main conclusions can be summarised as follows:
\begin{itemize}
\item   For stars with $\teff \ga$ 35 kK the flux in the \heii\ 
	continuum is increased 
	by 2 to 3 orders of magnitudes compared to predictions from 
	\nlte\ plane parallel model atmospheres (cf.~Gabler \etal 1989).
	With respect to Kurucz LTE models there is a 3-6 orders of
	magnitude increase at \teff\ $\ga$ 38000 K.

\item   The flux in the \hei\ continuum is not only increased due
	\nlte\ effects (e.g.~Kudritzki \etal 1991) but is also
	modified by wind effects (cf.~Najarro \etal 1996, Paper II).
	The combined effect of the mass outflow and \lb\ 
	lead to a {\em flatter energy distribution} in the 
	\hei\ continuum (see also Sellmaier \etal 1996).
	Typically our models lead to an increase of
	a factor of 1.25 to 2 for the \hei\ ionizing photon flux
	with respect to Kurucz models.
	
\item The Lyman continuum fluxes are modified due to \lb\ and stellar
	winds, although to a lesser degree than the spectrum at higher
	energies.
	For most cases the differences with Kurucz models
        are less than $\sim$ 20 \%  in the ionizing photon flux.
\end{itemize}

Using our calculations we provide revised ionizing fluxes for O3 to 
B0 stars (see Sect.~\ref{s_revision}) based on the recent temperature 
and gravity calibrations of Vacca \etal (1996).
The total number of Lyman continuum photons is found to be
slightly lower than those of Vacca et al., which are derived from
the plane parallel LTE models of Kurucz (1991).
Due to the increased flux in the \hei\ continuum, the hardness ratio 
$q_1/q_0$ of the \hei\ to H~{\sc i} continuum is increased by factors of
$\sim$ 1.6 to $\sim$ 2.5 depending on spectral type and luminosity class. 
These high hardness ratios only follow from Kurucz models at temperatures
of about 1000 to 11000 kK higher than used in our models.

We have discussed the assumptions inherent in our models
and point out future improvements. These improvements will be especially
important for understanding the EUV spectra of B-type stars, which
show relatively weak stellar winds. We have analysed the
EUV spectra of 
the recently observed B giants $\epsilon$ and $\beta$ CMa 
(Cassinelli \etal 1995, 1996) and have shown that reliable calculations
of the temperature structure in a model accounting for the
stellar wind and which includes a detailed treatment of photospheric blanketing
will be crucial to reproduce the EUVE observations. 
We argue that these effects are likely more important than
the pure wind effect invoked by Najarro \etal (1996) to explain
the EUV excess of $\epsilon$ CMa (Sect.~\ref{s_euve}).

Although potential shortcomings have been identified
(Sect.~\ref{s_improve}), we consider our predictions to be
fairly reliable for O stars (see also Schaerer 1996).
The recent study of Sellmaier \etal (1996) shows that \hii\ regions 
can provide very sensitive tests and that their exploratory models,
which are similar to the present ones, are quite successful in
explaining the observations.
To explore the broader impact of our new ionizing fluxes a grid of 
\hii\ region models has been calculated using the nebular
photoionization code {\em PHOTO} of Stasi\'{n}ska (see Stasi\'{n}ska
\& Schaerer 1996, Leitherer \etal 1996).

The spectral energy distributions are available
on request from the authors and will be included in a forth-coming
AAS CD-ROM (Leitherer \etal 1996).

%%%%%%%%%%%%%%%%%%%%%%%%%%%%%%%%%%%%%%%%%%%%%%%%%%%%%%%%%%%%%%%%%%%%%%%%
\acknowledgements{DS particularly thanks Andr\'e Maeder for his 
encouragement and support for this project.
Dietmar Kunze kindly provided results from his atmosphere calculations.
EUVE observations were made available by David Cohen.
DS also thanks Bill Vacca, Werner Schmutz and Jacques Babel
for numerous discussions and comments. 
This work was supported in part by the Swiss National Foundation of Scientific 
Research and by NASA through contract NAS5-31842.}

%\begin{figure}[htb]
%\centerline{\psfig{figure=figs/costar_7bis.eps,height=8.8cm}}
%\caption{}
%\label{fig_costar7bis}
%\end{figure}

%------------------------------------------------
% CoStar main sequence paper (III. Z=0.02 + 0.004)
%------------------------------------------------

\end{document}